\documentclass[aip,jcp,preprint,footinbib,floatfix,10pt]{revtex4-1}
\usepackage[utf8]{inputenc}

\usepackage{amsmath}%
\usepackage{graphicx}%
\usepackage{dcolumn}%
\usepackage{bm}%
\usepackage[caption=false]{subfig}%
\usepackage{float}%
\usepackage{multirow}%
\usepackage{verbatim}%
\usepackage{hyperref}%
\usepackage{xcolor}%
\usepackage{setspace}

\hypersetup{
 pdftitle={State-resolved coarse-grain cross sections for rovibrational excitation and dissociation of nitrogen based on ab initio data for the \texorpdfstring{$\mathbf{N_2}$-$\mathbf{N}$}{N2-N} system},
 pdfauthor=Erik Torres
}

\begin{document}

\title{State-resolved coarse-grain cross sections for rovibrational excitation and dissociation of nitrogen based on ab initio data for the \texorpdfstring{$\mathbf{N_2}$-$\mathbf{N}$}{N2-N} system}

\date{\today}

\author{Erik \surname{Torres}}
\email[Corresponding author: ]{torres@vki.ac.be}
\affiliation{Aeronautics and Aerospace Department, von Karman Institute for Fluid Dynamics, Chauss\'ee de Waterloo 72, 1640 Rhode-Saint-Gen\`ese, Belgium}
\author{Richard L. \surname{Jaffe}}
\affiliation{NASA Ames Research Center, Moffett Field, CA 94035, USA}
\author{David \surname{Schwenke}}
\affiliation{NASA Ames Research Center, Moffett Field, CA 94035, USA}
\author{Thierry E. \surname{Magin}}
\affiliation{Aeronautics and Aerospace Department, von Karman Institute for Fluid Dynamics, Chauss\'ee de Waterloo 72, 1640 Rhode-Saint-Gen\`ese, Belgium}

\begin{abstract} 
 In this paper, we present a method to generate state-resolved reaction cross sections in analytical form for rovibrational energy excitation and dissociation of a molecular gas. The method is applied to an \emph{ab initio} database for the $\mathrm{N_2}$-$\mathrm{N}$ system developed at NASA Ames Research Center. The detailed information on $\mathrm{N_2}$+$\mathrm{N}$ collisions contained in this database has been reduced by adapting a Uniform RoVibrational-Collisional bin model originally developed for rate coefficients.  Using a 10-bin system as an example, a comparison is made between two sets of coarse-grain cross sections, obtained by \emph{analytical inversion} and \emph{direct binning} respectively. The analytical inversion approach is especially powerful, because it manages to compress the entire set of rovibrational-level-specific data from the Ames database into a much smaller set of numerical parameters, sufficient to reconstruct all coarse-grain cross sections for any particular $\mathrm{N_2}$+$\mathrm{N}$-collision pair. As a result of this approach, the computational cost in in large-scale Direct Simulation Monte Carlo (DSMC) flow simulations is significantly reduced, both in terms of memory requirements and execution time. The intended application is the simulation of high-temperature gas-dynamics phenomena in shock-heated flows via the DSMC method. Such conditions are typically encountered in high-altitude, high-speed atmospheric entry, or in shock-tube experiments. Using this coarse-grain model together with ab initio rate data will enable more accurate modeling nonequilibrium phenomena, such as vibrationally-favored dissociation, an effect that is not well-captured by the conventional models prevalent in DSMC (i.e. Larsen-Borgnakke and Total Collision Energy). 
\end{abstract}

\pacs{82.20.-w, 47.70.Nd, 47.70.Fw, 47.40.Ki}

\maketitle


\section{Introduction} \label{sec:introduction}

The accurate modeling of thermo-chemical nonequilibrium effects in atmospheric entry flows using conventional Computational Fluid Dynamics (CFD), or particle-based Direct Simulation Monte Carlo~\cite{bird94a} (DSMC), requires reliable chemical-kinetic data for predicting the rates of internal energy excitation and molecular dissociation. For many years, computational fluid dynamicists had to rely on semi-empirical correlations, such as the ones by Millikan and White~\cite{millikan63a}, Appleton~\cite{appleton68a} and Park~\cite{park93a} to model these effects. Many of these data were originally obtained from experiments in chemical reactors, or shock tubes operating at much lower temperatures than those typical of atmospheric entry flows. By necessity, these data have routinely been extrapolated beyond their range of validity, and as a consequence even today most numerical simulations are affected by large uncertainties in the physico-chemical models. Often they fail to accurately predict the conditions in high-enthalpy ground tests, or flight measurement data. Since it is impractical, or downright impossible to fully reproduce the high-temperature conditions of atmospheric entry in ground-test facilities, the needed chemical-kinetic rate data can not be obtained by experimental methods alone. Fortunately, as computational power has increased over the last two decades, it has finally become feasible to apply the methods of computational chemistry to generate high-fidelity reaction rate data for the most relevant molecular systems in Earth's atmosphere. These data are based entirely on first-principles, or \emph{ab initio} quantum-chemistry calculations. Nevertheless, even today this remains a computationally expensive task, and for many mixtures of interest (i.e. Earth, or Mars atmosphere) full sets of cross-section data are not yet available. Thus, physics-based chemical rate data have been limited mostly to relatively simple systems, where only collisions between a single diatomic molecular species and its constituting atom(s) are taken into account. 

In this paper we focus on a gas mixture consisting exclusively of molecular- and atomic nitrogen. We will concentrate on $\mathrm{N_2}\left(v,J\right)$+$\mathrm{N}$ collisions, where $v$ and $J$ represent the discrete vibrational and rotational quantum numbers of molecular nitrogen in its ground electronic state. In the simulation of flows in thermo-chemical nonequilibrium, the state-to-state approach allows for the most detailed description of the gas mixture. It is more general than classical multi-temperature approaches~\cite{park90a} often followed in CFD modeling, since no \emph{a priori} assumptions in the populations of individual rovibrational energy levels are required and non-Boltzmann internal energy distributions may emerge naturally as a result of the flow conditions being simulated. In order to extract the greatest amount of detail from quantum-chemistry-derived reaction rate data, we will therefore resort to a state-to-state description in our CFD/DSMC codes.

Over the last two decades, several groups worked to introduce \emph{ab initio} reaction rate data into gas-kinetic simulations methods and apply them to hypersonic nonequilibrium flows. Bruno et al~\cite{bruno02a} first implemented a vibrationally-specific model for nitrogen in a DSMC code. In their work they resolved 66 discrete vibrational levels of the $\mathrm{N_2}$-molecule. For diatom-atom collisions, i.e. $\mathrm{N_2} \left( v \right) + \mathrm{N}$, they used cross-section data based on quasi-classical trajectory (QCT) calculations performed by the group at the University of Bari. Data were available for both excitation/deexcitation~\cite{esposito00a} and dissociation~\cite{esposito99a}. This was one of the earliest instances where QCT databases were used successfully within DSMC. More recently, Li and Levin~\cite{li12a} have combined cross sections based on the forced harmonic oscillator (FHO) model for vibrational-translational (VT) transitions and on QCT calculations for dissociation reactions in state-to-state DSMC simulations of shock waves in nitrogen. One very detailed set of cross-section data is based on the \emph{ab initio} QCT calculations of Jaffe et al.~\cite{jaffe08a} for the $\mathrm{N_2(X^1\Sigma_g^+)} + \mathrm{N({}^4S_u)}$-system (often referred to as ``N3''), originally compiled at NASA Ames Research Center. The NASA Ames database is comprised of state-resolved collision cross sections and thermal rate coefficients for $\mathrm{N_2}$+$\mathrm{N}$ collisions across the 9390 ro-vibrational levels of the ground electronic state. Cross sections and thermal rate coefficients for inelastic processes (i.e., ro-vibrational energy transfer) and exchange and dissociation processes are tabulated. This database has been used extensively in CFD studies of the dynamics of internal energy excitation and dissociation in nitrogen mixtures by Magin et al~\cite{magin10a, panesi13a}, while Kim and Boyd~\cite{kim12a} have integrated it into a DSMC solver. More recently, Parsons et al~\cite{parsons14a} generated state-resolved (using a coarse-grained model for lumping together rovibrational levels) dissociation cross sections for $\mathrm{N_2} + \mathrm{N_2}$-collisions also using the QCT method. However, their calculations made use of a new global \emph{ab initio} potential energy surface (PES) valid for both $\mathrm{N_2}$+$\mathrm{N_2}$ and $\mathrm{N_2}$+$\mathrm{N}$ interactions, generated by the Truhlar group~\cite{paukku13a, paukku14a}. This same PES was used in subsequent works to generate cross sections for rovibrational excitation by Li et al~\cite{li15a, zhu16a}. The Minnesota N4 PES was further improved and used by Bender et al~\cite{bender15a} to generate multi-temperature dissociation rate coefficients for $\mathrm{N_2}$+$\mathrm{N_2}$-collisions.

Both hydrodynamic and kinetic simulations using the full set of levels are extremely computationally expensive, due to the large number of internal energy states and processes involved. Thus, they have mostly been limited to master equation studies involving space-homogeneous heat baths~\cite{panesi13a} and one-dimensional inviscid flows, such as behind normal shock waves~\cite{panesi14a}. One way to remedy this situation has been to propose so-called coarse-grain, or reduced-order state-to-state models. The details of the reduction vary, but the basic concept is always to approximate the behavior of the full set of levels with a much smaller number of suitably defined internal energy classes, whose properties are weighted averages over the properties of the individual constituting levels. This lumping-together of internal energy states automatically leads to a significant reduction in the number of associated state-to-state reaction rates and greatly reduces the cost of simulations. Here we enumerate the most important coarse-graining strategies that have been proposed in recent years, with particular emphasis on those used with the N3 and N4-systems in mind.

A vibrational-specific collisional (VC) model was originally used by Bruno and co-workers~\cite{bruno02a} to simulate thermochemical nonequilibrium in shock waves using DSMC. In their implementation the roughly 60 vibrational levels of the $\mathrm{N_2}$-molecule were treated using a state-to-state approach, while the rotational energy exchange was handled using the conventional Larsen-Borgnakke model~\cite{borgnakke75a}. More recently, within the context of classical CFD, Munaf\`o et al.~\cite{munafo12a} used a vibrational-collisional model to study the chemical nonequilibrium and chemically frozen flow in nozzle expansions. In this work, the independently obtained QCT databases of NASA Ames~\cite{jaffe08a} and of the Bari group~\cite{esposito99a, esposito00a} where compared. The VC model maintains the classic paradigm of fully decoupling the vibrational and rotational energy distributions. Since at low temperatures rotational-translational (RT) relaxation is known to proceed much quicker than vibrational-translational (VT) relaxation, in the hydrodynamic description the VC model is formulated so as to assume that the rotational levels always follow a Boltzmann distribution at the common rotational-translational temperature $T = T_\mathrm{rot}$. This equilibrium assumption was found in later work~\cite{panesi14a} to be the cause for overestimation of the dissociation rates within shock waves. Any coarse-grain model, that is supposed to overcome this deficiency should therefore be formulated without this assumption.
 
An early attempt at formulating such a model was the aforementioned Uniform Rovibrational collisional (URVC) bin model. It has been studied by Magin et al~\cite{magin12a} in the context of classical CFD applications, and forms the basis for the work discussed in the present paper. In the URVC model, all rovibrational levels of the $\mathrm{N_2}$-molecule are lumped together into a small number of energy ``bins''. Unlike in the VC model, the levels to be binned are not selected according to their vibrational quantum numbers, but rather according to their overall rovibrational energy. This means that a single bin may contain a mix of levels with a wide range of rotational and vibrational quantum numbers, but with similar energy. As a consequence, it is sometimes referred to as an ``energy-based'' binning approach. The binning procedure for the URVC model is reviewed in Sec.~\ref{sec:urvc_binning}. 

Beyond the URVC model, the Boltzmann Rovibrational collisional (BRVC) bin model has been particularly successful in reproducing the thermodynamic properties and chemical-kinetic behavior of the full set of levels~\cite{munafo14c, munafo14d}. This model and a later evolution, the ``Hybrid'' bin vibrational collisional (HyBVC)~\cite{munafo15a} model are able to accurately resolve individual level populations within each bin using a relatively small number of bins. Macdonald et al~\cite{macdonald18b} have recently used the BRVC and VC bin models to directly obtain coarse-grained rate coefficients for the more complex $\mathrm{N_2}(X^1\Sigma_g^+)-\mathrm{N_2}(X^1\Sigma_g^+)$ system, based on QCT calculations on the NASA Ames N4 PES~\cite{jaffe10a}. Unlike for N3, a full rovibrational state-resolved approach is impossible for this system (at least for the foreseeable future), since it would require on the order of $10^{1 \rightarrow 5}$ reaction rate coefficients (or cross sections) to be computed to include all detailed reaction paths. At present the only known alternative for overcoming this problem is to perform QCT calculations on the \emph{ab initio} PES ``on the fly'' as part of a DSMC simulation, replacing any stochastic collision routines, which would require a full cross section database. This method, referred to as direct molecular simulation (DMS) has recently been developed by Schwartzenruber's group~\cite{schwartzentruber17a}. In this method, all the information needed to determine the outcome of a collision is provided by the relevant \emph{ab initio} PES. While the random selection of collision pairs in a given cell is done following the usual conventions of DSMC (e.g. using Bird's NTC scheme and a suitable hard sphere total cross section~\cite{bird94a}), classical trajectory calculations are then performed for each interaction involving molecule-molecule, or molecule-atom interactions. Their approach was originally based on Koura's classical trajectory calculation (CTC-DSMC) method~\cite{koura97a, koura98b, koura02a}. Early versions of Schwartzenruber's method~\cite{norman13a, valentini14b} used simple potentials assuming fixed intra-molecular bond lengths during $\mathrm{N_2}$+$\mathrm{N_2}$ interactions. This prevented the method from simulating dissociation. However, more recently these empirical potentials have been replaced by high-fidelity \emph{ab initio} PES's from Truhlar's group for simulating rovibrationally coupled excitation and dissociation in nitrogen~\cite{valentini15a, valentini16a} and oxygen~\cite{grover18a} in isothermal heat baths. The DMS method has also been used in conjunction with the NASA Ames N3~\cite{grover17b} and N4 potentials~\cite{macdonald18c}. In the latter case, DMS and the coarse-grained master equation approach of Macdonald et al~\cite{macdonald18b} show very good agreement.

In another recent paper, Sahai et al~\cite{sahai17a} discuss two alternative strategies (so-called \emph{island} and \emph{spectral} clustering), which abandon the ``energy-based'' binning approach in favor of more complex criteria for selecting levels to be grouped together. By basing the selection process on the level-to-level relaxation rates in addition to the difference between level energies themselves, they aim to group together rovibrational levels which ``equilibrate quickly'' among each other into a common bin, while keeping levels connected via ``slowly relaxing'' processes in separate bins. The majority of these coarse-grained models require use of either the translational mode- or suitably defined internal energy mode temperatures to determine the level populations inside the bin. This poses no problem in classical CFD, which is based on a hydrodynamic flow description and where the temperature forms part of the set of solution variables. Unfortunately, the same approach cannot be easily extended to the DSMC method, which simulates the gas at the kinetic scale. In such a kinetic flow description  temperatures only appear as macroscopic moments, representing the collective energy distribution over an ensemble of particles. Thus, using a macroscopic temperature to specify the internal energy state of individual DSMC particles in the simulation algorithm is inconsistent with the formulation and may lead to problems in obtaining physically sound DSMC solutions~\cite{torres13a}. Therefore, the BRVC, HyBVC and related models will not be used in this paper and the subsequent DSMC implementation.

Another interesting alternative to the energy-based binning~\cite{magin12a} has been pursued by Zhu et al~\cite{zhu16a}. They have proposed a so-called two-dimensional binning strategy, where the choice of levels to be lumped together does not depend on the overall energy, but instead on the relative proximity of their particular vibrational and rotational quantum numbers. Using this approach, and after additional refinement of the lowest-energy bins they obtained a final system consisting of 99 bins. They then proceeded to generate bin-specific cross sections for rovibrational energy exchange in $\mathrm{N_2} - \mathrm{N}$ collisions, derived from two distinct sets of QCT data (based on the NASA Ames~\cite{jaffe08a}- and Truhlar~\cite{paukku13a} PES's for N3 respectively). These cross sections were then used in a DSMC study of rovibrational relaxation in nitrogen, within an isothermal reservoir. Although more involved than the energy-based URVC binning used in our paper, this 2D-binning approach may prove advantageous when the precise populations of rotational and vibrational levels have to be reconstructed from a bin-based flow solution (e.g. for determining radiative properties of the gas within the flow field, which are especially sensitive to the populations of specific rovibrational levels).

It should be noted that coarse-grained models have existed for much longer than the previous references may suggest. As far back as the late 1980's, Haug et al.~\cite{haug87a, haug92a} reduced a set of rovibrational cross sections for hydrogen-argon collisions using vibrational-level-specific, rotational-level-specific, as well as other custom binning combinations. Although the hydrogen molecule possesses significantly fewer rovibrational levels than nitrogen (162 in Haug's case vs. $\approx$9400 in our case), in both cases the coarse-graining approach is very useful in reducing the complexity of the original system while maintaining the main features required for describing nonequilibrium reaction rates.

The main goal of this paper is to present a simple, easy-to-use method by which to generate a set of state-to-state reaction cross sections for rovibrational excitation and dissociation of molecules, originating from \emph{ab initio} data. These data-sets can be used at a later stage together with the DSMC method to simulate a variety of rarefied flows, such as those encountered in high-speed, high-altitude atmospheric entry. For this purpose, we will first perform a brief analysis of the NASA Ames N3 database, our source of quantum-chemistry data. Given that the Ames database contains information on such a vast number of detailed chemistry processes, it is impractical to integrate it one-to-one into a DSMC code intended for large-scale flow simulations. Therefore, we will also apply coarse-graining techniques to reduce the size of the final cross section data-sets. As mentioned earlier, several model reduction techniques exist, but most of them are only directly applicable in the context of classical CFD~\cite{munafo14c, munafo15a, sahai17a}. Alternative reductions, more suitable for DSMC have been proposed recently~\cite{zhu16a}. Nevertheless, here we will continue to follow the original energy-based approach of the URVC bin model, because we believe that its simplicity provides a number of advantages when used in the context of the DSMC method.

With a 10-URVC-bin system used as an example, we then show how to extract state-resolved cross sections for internal energy exchange and molecular dissociation from the database via two different methods. The first is a straightforward \emph{direct binning} of all the rovibrational-level-specific cross sections. This will result in a set of tabulated, collision-energy-dependent bin-specific cross sections. The second method consists in extracting equivalent cross sections by means of \emph{analytical inversion} of the corresponding rate coefficients. This approach will yield a set of bin-specific parameters, which will allow us to specify the precise ``shape'' of each cross section using a simple analytical form. We will then weigh the relative advantages and drawbacks of the two cross-section sets and select a preferred approach for further use with DSMC. 

In order to make the URVC bin model fully consistent with the DSMC method, we will propose several changes to its original formulation~\cite{magin12a}. As a consequence of these changes we will no longer attempt to enforce detailed balance relations at the resolution of individual rovibrational levels. Instead, we formulate the forward- and reverse macroscopic rate coefficients and reaction cross sections in a manner so as to ensure detailed balance only for the bin-averaged populations, which then completely replace the original level-resolved ones. As a consequence, the thermodynamic- and chemical-kinetic properties of the gas mixture are then exclusively governed by the bin populations. We intend to set up the URVC bin model in a manner simple enough to make its integration within the framework of a DSMC solver as practical as possible. At the same time, these changes ensure that we will be able maintain consistency with the macroscopic balance equations valid at the hydrodynamic scale.

This paper is structured as follows. Sec.~\ref{sec:nasa_n3} provides a short overview of the quantum-chemical database for $\mathrm{N_2}\left(v,J\right)$+$\mathrm{N}$-collisions developed at NASA Ames Research Center. Next, in Sec.~\ref{sec:urvc_binning} the uniform rovibrational collisional (URVC) bin model for reducing the N3 data-set is reviewed, along with the recent modifications to the model. Following this, Sec.~\ref{sec:cross_section_extraction} discusses the two alternative methods for extracting coarse-grained cross-section data from the Ames database. Finally, Sec.~\ref{sec:conclusions} contains the conclusions and an outlook on the future work.


\section{NASA Ames database} \label{sec:nasa_n3}

It should be noted that the NASA Ames N3 database discussed in this paper is not the only set of QCT-based cross sections for nitrogen currently available. In the late 1990's Capitelli's group published results on vibrational-specific excitation and dissociation rates for the N3-system, with rotational level populations sampled from a Boltzmann distribution at $T_\mathrm{rot} = T$~\cite{esposito99a, esposito00a}. More recently, the Truhlar group has published results on a potential energy surface for the N4-system~\cite{paukku13a, paukku14a}. Their PES has been used to perform trajectory calculations to determine multi-temperature dissociation rate coefficients for the reactions $\mathrm{N_2} + \mathrm{N_2} \rightarrow \mathrm{N_2} + 2 \, \mathrm{N}$ and $\mathrm{N_2} + \mathrm{N_2} \rightarrow 4 \, \mathrm{N}$~\cite{bender15a}. The NASA Ames database~\cite{jaffe08a, chaban08a, schwenke08a, jaffe09a} discussed here, contains a detailed set of kinetic rate data to describe the inelastic and reactive interactions of $\mathrm{N_2}$-$\mathrm{N}$ and $\mathrm{N_2}$-$\mathrm{N_2}$ pairs. Although the present work focuses only on the study of  $\mathrm{N_2}-\mathrm{N}$ collisional processes, the study of the $\mathrm{N_2}$-$\mathrm{N_2}$ dynamics is ongoing~\cite{jaffe10a, panesi13b}. 

\subsection{Rovibrational energy levels of \texorpdfstring{$\boldsymbol{\mathrm{N_2}}$}{N2}} \label{sec:rovibrational_levels}

\begin{figure}
 \centering
 \includegraphics[width=\columnwidth]{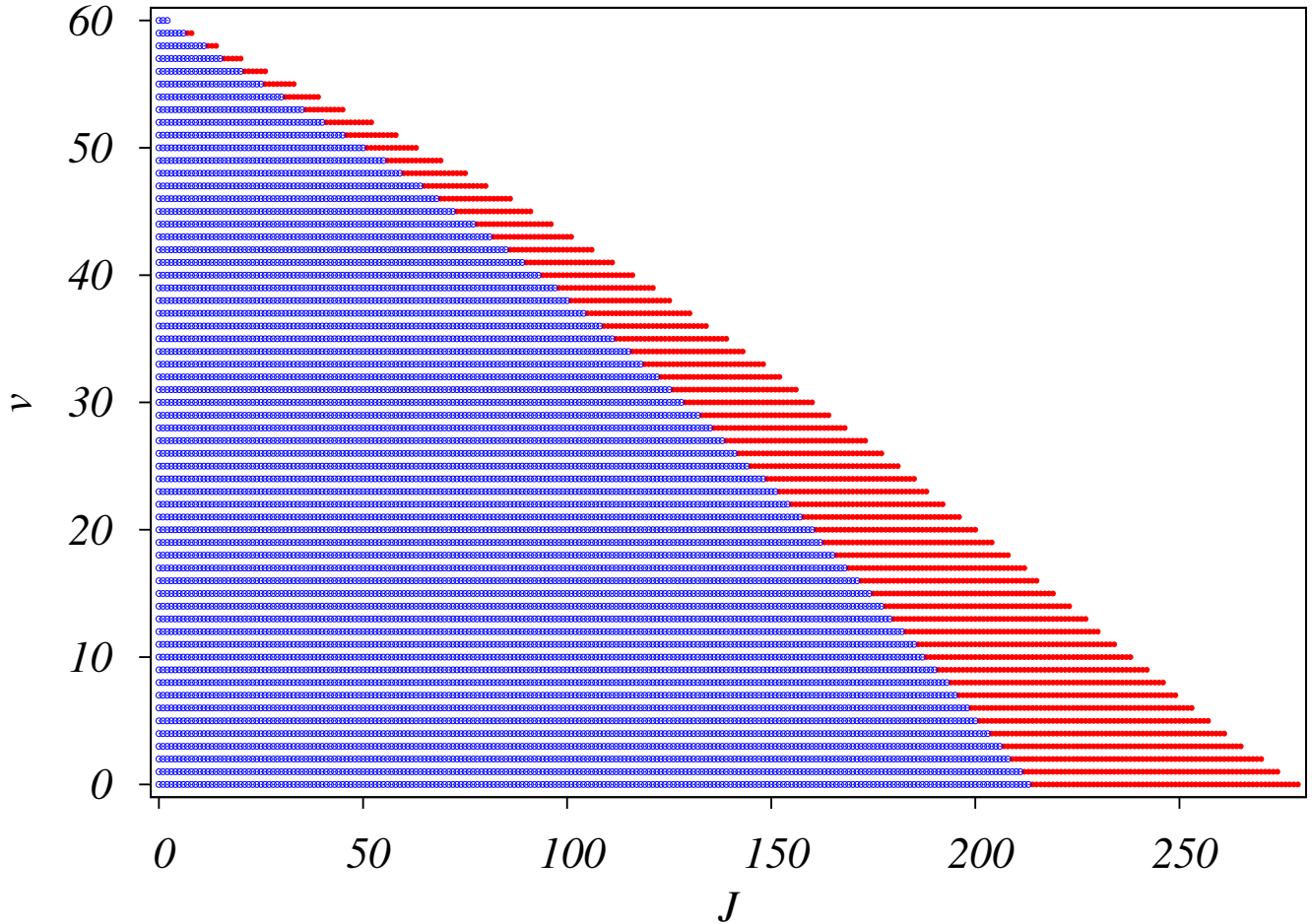}
 \caption{Distribution of rovibrational levels of $\mathrm{N_2}$ sorted by rotational and vibrational quantum numbers: $\left( J, v \right)$: Blue open circles designate bound, filled circles in red represent quasi-bound levels.}
 \label{fig:vj_map}
\end{figure}

The energies of the 9390 rovibrational levels in the ground electronic state of $\mathrm{N_2}$ were determined using quantum mechanics calculations within the WKB approximation~\cite{schwenke88a} using a modified version of the high-fidelity potential of Le Roy et al.~\cite{leroy06a} Each level is identified by a unique pair of vibrational and rotational quantum numbers $\left( v, J \right)$, comprised within the ranges $v = 0, \ldots, 60$ and $J = 0, \ldots, 279$. When all levels are arranged in increasing order according to their particular energies~\footnote{This choice is arbitrary, but it is convenient for energy-based binning, which is considered later.}, it is possible to map their quantum number pairs $\left( v, J \right)$ to a global index $i$. The two notations are related by:
\begin{equation}
   i = i\left(v, J\right), \; v = 0, \ldots, v_\mathrm{max}, \; J = 0, \ldots, J_\mathrm{max}\left(v\right), \label{eq:vj_forward_relation}
\end{equation}
with the inverse relation:
\begin{equation}
 v = v \left( i \right), \quad J = J \left( i \right), \quad i \in \mathcal{I}_\mathrm{BP}, \label{eq:vj_inverse_relation}
\end{equation}
where $\mathcal{I}_\mathrm{BP}$ represents the global set of level indices, comprising both truly bound and ``pre-dissociated'' (also known as quasi-bound) levels. Their distribution over all valid $\left( v, J \right)$-combinations is sketched in Fig.~\ref{fig:vj_map}. 

\begin{figure}
 \includegraphics[width=\columnwidth]{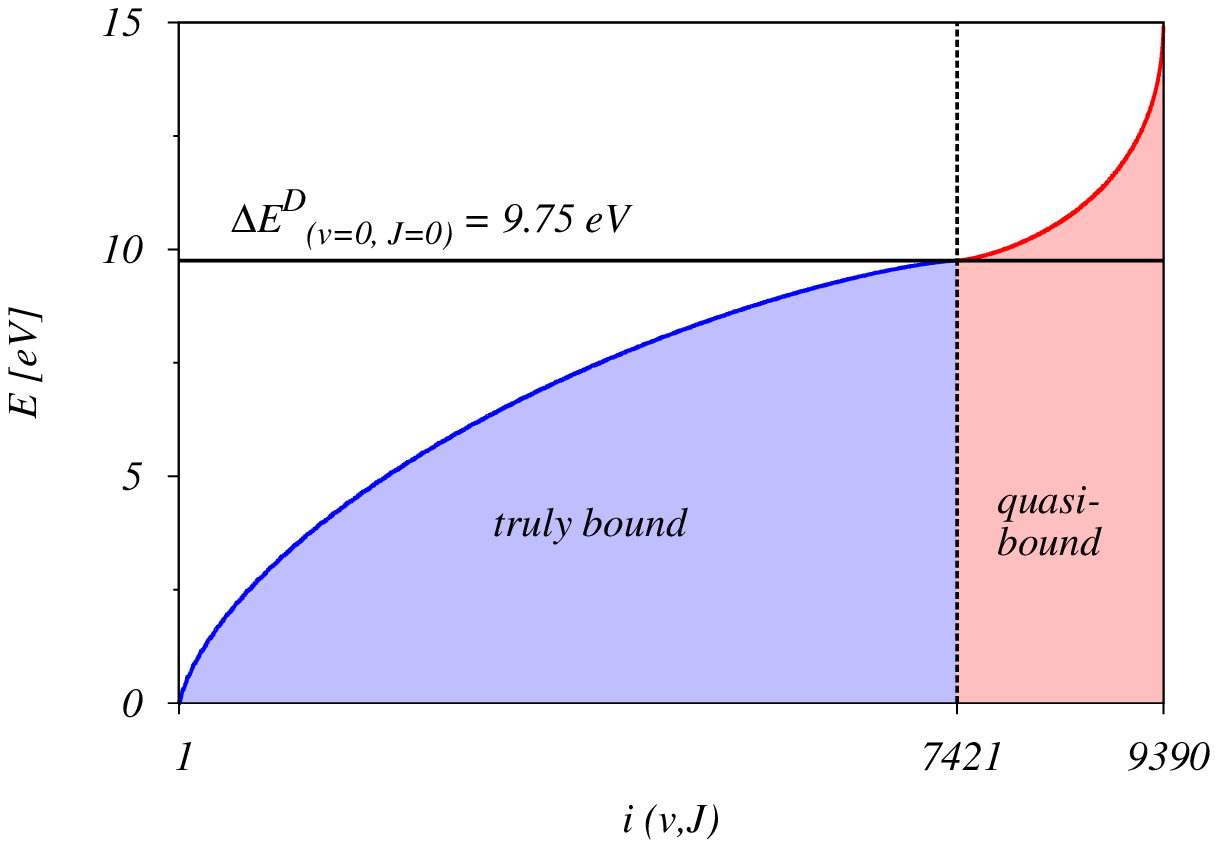}
 \caption{Discrete rovibrational energies (ordinate) sorted in ascending order as a function of level index $i \left(v,J\right)$ (abscissa). Levels with energy lower than $9.75 \, \mathrm{eV}$ are truly bound (blue region). Those with higher energies are quasi-bound or pre-dissociated. (red region). The highest-energy truly bound level has index $7421$.}
 \label{fig:rovibrational_levels_energy}
\end{figure}

Truly bound (or just ``bound'') levels are those whose energy lies below the threshold $\Delta E_{\left(v=0, J=0\right)}^D = 9.75 \, \mathrm{eV}$, which is the exact amount of energy required to dissociate a nitrogen molecule from its rovibrational ground state. It corresponds to the difference between the formation energies of the two newly created nitrogen atoms and that of the original molecule: $\Delta E_{\left(v=0, J=0\right)}^D = 2 \, E_\mathrm{N} - E_{\mathrm{N_2} \left(v=0, J=0\right)}$\footnote{By convention, the energy of formation of the molecule in its ground rovibrational state has been assumed to be zero: $E_{\mathrm{N_2} \left(v=0, J=0\right)} = 0$.}. Any nitrogen molecule populating a bound level will possess an amount of rovibrational energy, which lies in the range $0 < E_{\left( v, J \right)} < \Delta E_{\left(v=0, J=0\right)}^D$. These energy levels are shown in Fig.~\ref{fig:rovibrational_levels_energy}, left of the vertical dividing line. By contrast, molecules which populate quasi-bound levels have internal energies lying above $\Delta E_{\left(v=0, J=0\right)}^D$ and are located to the right of the dividing line in Fig.~\ref{fig:rovibrational_levels_energy}. Spontaneous dissociation of such pre-dissociated molecules, \emph{without} interaction with another particle, occurs due to quantum-tunneling. Collision-induced dissociation from pre-dissociated levels, on the other hand, proceeds in much the same manner as from bound levels, and contributes to a much greater degree to the overall dissociation rate~\cite{panesi13a}.

Based on this sub-division, one may define the set of bound levels $\mathcal{I}_\mathrm{B} = \left\lbrace 1, \ldots, 7421 \right\rbrace$, as well as the set of all pre-dissociated levels $\mathcal{I}_\mathrm{P} = \left\lbrace 7422, \ldots, 9390 \right\rbrace$. Of course, one has that $\mathcal{I}_\mathrm{BP} = \mathcal{I}_\mathrm{B} \cup \mathcal{I}_\mathrm{P}$ and $\mathcal{I}_\mathrm{B} \cap \mathcal{I}_\mathrm{P} = \emptyset$. Figs.~\ref{fig:vj_map} and \ref{fig:rovibrational_levels_energy} make it clear that the majority of rovibrational levels of $\mathrm{N_2}$ are bound. The degeneracy associated with each rovibrational level is a function of the rotational quantum number: $g_i = \left( 2 \, J\left(i\right) + 1 \right) \, g_i^\mathrm{NS}$, where the nuclear spin degeneracy is $g_i^\mathrm{NS} = 6$ for even-valued $J\left(i\right)$ and $g_i^\mathrm{NS} = 3$ for odd-valued $J\left(i\right)$. The degeneracy for the nitrogen atom is $g_\mathrm{N} = 12$ and includes nuclear- and electronic spin contributions.


\subsection{Rovibrational level-specific cross sections} \label{sec:level_specific_processes}

The Ames N3 database comprises a listing of level-specific cross sections for more than $20 \times 10^6$ detailed processes~\cite{panesi13a} involving $\mathrm{N_2}+\mathrm{N}$-collisions. They can be classified into two main types:
1) Collisional dissociation from bound- and pre-dissociated states
\begin{equation}
 \mathrm{N_2} \left( i \right) + \mathrm{N} \overset{\sigma_i^D}{\rightarrow} \mathrm{N} + \mathrm{N} + \mathrm{N}, \quad i \in \mathcal{I}_\mathrm{BP}, \label{eq:dissociation_levels}
\end{equation}

2) Collisional transitions between arbitrary rovibrational levels
\begin{equation}
 \mathrm{N_2} \left( i \right) + \mathrm{N} \overset{\sigma_{i \rightarrow j}^E}{\rightarrow} \mathrm{N_2} \left( j \right) + \mathrm{N} \quad i, j \in \mathcal{I}_\mathrm{BP}, \quad \label{eq:nonreactive_collisions_levels}
\end{equation}

Comprised within the definition of Eq.~(\ref{eq:nonreactive_collisions_levels}) are a) \emph{excitation} reactions for which $E_j > E_i$, b) \emph{deexcitation} reactions where $E_j < E_i$ and c) \emph{elastic} collisions when $E_j = E_i$. Integrated cross sections over all deflection angles $\sigma_{i \rightarrow j}^E$ for all three sub-categories, as well as the dissociation cross sections $\sigma_i^D$, are tabulated within the Ames database as a function of the pre-collision relative translational energy $E_t$. In the definition of Eq.~(\ref{eq:nonreactive_collisions_levels}) we have tacitly included so-called \emph{exchange} reactions, where one of the nitrogen atoms is swapped between the two collision partners, e.g. $\mathrm{N}^a\mathrm{N}^b \left( i\right) + \mathrm{N}^c \rightarrow \mathrm{N}^a + \mathrm{N}^b\mathrm{N}^c \left( j\right)$. It is known that such atom-exchange reactions contribute significantly to the overall rate of internal energy exchange rate~\cite{panesi13a} and cannot be neglected. However, in the particular case of the N3-system the distinction between regular excitation and exchange reactions is only apparent at the QCT-calculation stage, where each atom's individual trajectory can be identified at all times. From a macroscopic point of view though, both processes result in the same outcome, and thus their cross sections are added together. Pre-dissociation reactions, on the other hand, are ignored in this paper, as the process' influence on the dynamics of the system was found to be negligible by Panesi et al~\cite{panesi13a}.

Rovibrational state specific cross sections for the collisional processes comprised by Eq.~(\ref{eq:dissociation_levels}) and (\ref{eq:nonreactive_collisions_levels}) were originally generated as statistical averages over a large number of QCT calculations~\cite{truhlar79a, schwenke88a, esposito00a} involving three separate nitrogen nuclei (hence the informal designation ``N3''). A pre-requisite for such calculations is an accurate representation of the inter-atomic potential energy surface (PES), which allows to determine the forces exerted on the three nitrogen nuclei as they interact with each other during a collision. Determination of this PES is based on \emph{ab initio} quantum-mechanical electronic structure calculations. A comprehensive description of the methodology for the NASA Ames database is given by Jaffe et al.~\cite{jaffe15a}. Here we only recall that the maximum impact parameter in the trajectory calculations was 4.2~\AA, and the pre-collision energy was sampled from discrete values over the range 0.03 to 50.2 eV.

Due to the limited number of trajectories available for each individual transition, the cross sections are subject to statistical noise. The magnitude of this Monte Carlo statistical sampling error is approximately inversely proportional to the square root of the sample size. In some instances, certain state-to-state transitions, though physically possible, are not found in the database. Missing trajectories for a particular collision outcome point to a naturally low probability of occurrence of this particular process. In many other cases, both excitation cross sections $\sigma_{i \rightarrow j}^E$ and the corresponding deexcitation cross sections $\sigma_{j \rightarrow i}^E$ were present in the database. These two must verify micro-reversibility relations~\cite{giovangigli99a, nagnibeda09a}:
\begin{equation}
 g_i \, E_t \, \sigma_{i \rightarrow j}^E \left( E_t \right) = g_j \, E_t^\prime \, \sigma_{j \rightarrow i}^E \left( E_t^\prime \right), \qquad i,j \in \mathcal{I}_\mathrm{BP}, \label{eq:detailed_balance_levels}
\end{equation}
with the ``primed'' energy equal to $E_t^\prime = E_t + E_i - E_j$. However, due to the inherent statistical noise caused by the finite sample sizes used, the excitation- and deexcitation cross sections recorded in the database do not always exactly satisfy Eq.~(\ref{eq:detailed_balance_levels}) at all corresponding values of $E_t$  and $E_t^\prime$. This poses a problem, since it is not automatically clear how best to make use of the N3 database. If one were to use all entries in the database without modification, many of the processes for which forward and backward cross sections are listed would not satisfy Eq.~(\ref{eq:detailed_balance_levels}). This would make it impossible to retrieve the correct equilibrium level populations in actual master equation, or DSMC simulations. One remedy would be to only utilize the tabulated cross sections in one ``direction'', i.e. only use cross sections for endothermic, or conversely only for exothermic reactions. After selecting one option, one would then re-compute the corresponding reverse cross sections using Eq.~(\ref{eq:detailed_balance_levels}). This is what Jaffe et al~\cite{jaffe15a} originally proposed. They opted for retaining only the exothermic cross sections, since these were more frequent in the database by a factor of roughly 2:1. They argued that these cross section were also more accurate, because most trajectories resulted in exothermic reactions, reducing the Monte Carlo sampling error. Here, however we have opted for a slightly different approach. In order to make use of all entries in the database, all available deexcitation cross sections were first transformed via Eq.~(\ref{eq:detailed_balance_levels}) to their corresponding endothermic cross sections. Then, if entries for that particular process were already available in the database, an average of the two quantities was computed. In cases where the excitation cross sections for a given process were missing from the database, using detailed balance allowed us to fill in some of the gaps. The same operation was then also applied in the opposite sense.

\begin{figure}
 \centering
 \includegraphics[width=\columnwidth]{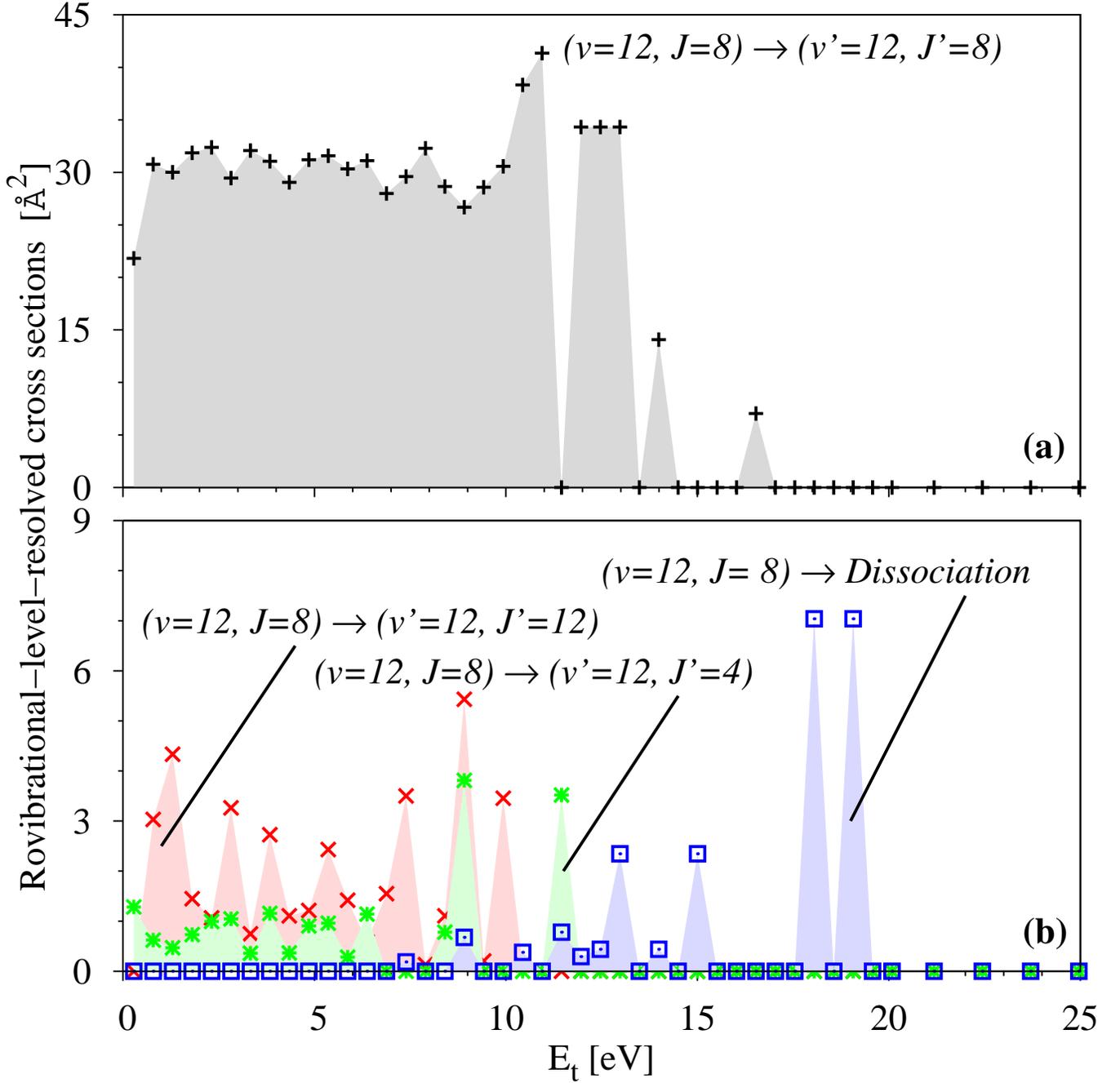}
 \caption{State-resolved total cross sections (integrated over all post-collision deflection angles) from the N3 database. Upper graph: elastic cross section for rovibrational level $(v = 12, J = 8)$. Lower graph: selected inelastic cross sections for rovibrational level $(v = 12, J = 8)$. Shaded in red: excitation to level $(v^\prime = 12, J^\prime = 12)$; shaded in green: deexcitation to level $(v^\prime = 12, J^\prime = 4)$; shaded in blue: dissociation.}
 \label{fig:sample_cross_sections}
\end{figure}

Four representative cross sections directly extracted from the Ames database are plotted in Fig.~\ref{fig:sample_cross_sections}. In the upper graph, Fig.~\ref{fig:sample_cross_sections}(a) shows the elastic cross section for level $i = 1000$, equivalent to $(v = 12, J = 8)$, as a function of $E_t$, i.e. the relative translational energy of the two collision partners. Several inelastic cross sections for the same pre-collision level are shown in Fig.~\ref{fig:sample_cross_sections}(b). For a given pre-collision level, the elastic cross section is typically several times larger than the inelastic ones. This can be attributed to the fact that most elastic collisions, especially at larger impact parameters, are ``grazing'' collisions, causing only small trajectory changes of the collision partners. In this case, the interaction between the two collision partners is often not strong enough to disturb the internal structure of the nitrogen molecule. The elastic cross section in Fig.~\ref{fig:sample_cross_sections}(a) is non-zero immediately above $E_t = 0$ and remains relatively constant around 30 $\text{\AA}^2$ until approximately $E_t = 10$ eV. At higher collision energies its magnitude begins to decrease, and beyond 15 eV the elastic cross section is practically zero. At such high relative collision speeds, the two particles have very little time to interact and collisions become less probable. 

At this point it must be noted that the elastic cross section shown here is, in a sense, flawed. During the original job of generating the N3 database, the values for $\sigma_{ii}^E$, all producing curves similar to the one in Fig.~\ref{fig:sample_cross_sections}(a), only appeared as a by-product of the actual goal of obtaining state-to-state QCT rate coefficients for inelastic collisions and dissociation. In order to converge the inelastic/dissociation cross sections (or more precisely the associated rate coefficients), quasi-classical trajectories were run for the range of impact parameters $0 \le b \le b_\mathrm{max} = 4.2$\AA~\cite{jaffe15a}. For every given pre-collision energy $E_t$ and impact parameter $b$, individual cross sections were then obtained as the product of a hard-sphere cross section $\pi b_\mathrm{max}^2$ and the probability of a specific state-to-state transition/dissociation, e.g. $\sigma_{i \rightarrow j}^E \left( E_t \right) = \pi b_\mathrm{max}^2 P_{i \rightarrow j}^{E, \mathrm{QCT}} \left( E_t \right)$. These probabilities, in turn, were obtained as the ratio of the number of trajectories producing the given transition and the total number of trajectories run at the particular collision energy, e.g. $P_{i \rightarrow j}^{E, \mathrm{QCT}} \left( E_t \right) = \left[ N_{i \rightarrow j}^E / N_\mathrm{tot} \right] \left( E_t \right)$. Any trajectories, which did not produce an internal state transition or dissociation, i.e. $N_{ii} = N_\mathrm{tot} - \sum_{j \in \mathcal{I}_\mathrm{BP}, j \ne i} N_{i \rightarrow j} - N_i^D$, were then ``left over'' to count toward the cross section $\sigma_{ii}^E$. By construction, the upper limit for $\sigma_{ii}^E$ would be a constant $\pi b_\mathrm{max}^2 = 55.42$\AA, precisely if none of the trajectories resulted in inelastic/dissociation collisions, regardless of the collision energy. This is, of course, only an arbitrary upper limit dictated by the value of $b_\mathrm{max}$. In general, classical mechanics will produce an infinite elastic cross section, unless $b_\mathrm{max}$ is fixed, or conversely an arbitrary lower bound for the deflection angle is introduced. Rigorously computing the elastic cross section can only be accomplished with a scattering theory based on quantum mechanics~\cite{jaffe15a}. However, such a task is beyond the scope of the present paper and we will continue to use the elastic cross sections from the database, despite of their limitations. In fact, our approach can be considered similar to the one of Kim and Boyd~\cite{kim14a}, who fitted their QCT-derived total cross sections to analytical expressions. They also had to impose a $b_\mathrm{max}$ in order to obtain finite values.

The excitation and deexcitation cross sections to levels $(v^\prime = 12, J^\prime = 12)$ and $(v^\prime = 12, J^\prime = 4)$, respectively, have sizable values up to around 10-12 eV, but decrease quickly at higher collision energies. Compared to the elastic cross section, they exhibit much stronger oscillations in magnitude. These oscillations may be a physical feature of this type of interaction, where transitions are much more probable at specific energy intervals, but may also be due to statistic noise introduced by the limited number of trajectories available to compute each cross section. In terms of energy difference, the pre- and  post-collision levels chosen here lie fairly close to one another. It is assumed that such transitions are much more likely than those involving larger energy jumps and that they contribute to most of the internal energy exchanged in the gas. The third cross section shown in Fig.~\ref{fig:sample_cross_sections}(b) is for dissociation from level $(v = 12, J = 8)$. The energy of this level lies approximately 3.25 eV above the rovibrational ground state. Therefore, the minimum energy required for dissociation is $\Delta E_{\left( v = 12, J = 8 \right)}^D = 2 \, E_N - 3.25 \, \mathrm{eV} = 6.5 \, \mathrm{eV}$, and below this threshold the dissociation cross section must be zero. Beyond this value, the dissociation cross section increases and oscillates slightly in the range of 1 to 2 $\text{\AA}^2$ up to around $E = 15$ eV. Finally, it exhibits two large peaks at energies around 17 and 19 eV, before it then drops down to zero at higher energies. Due to the large number of individual processes in the database, not all cross sections could be thoroughly examined. However, it is assumed that their behavior will be similar to that exhibited by the samples in Fig.~\ref{fig:sample_cross_sections}.

Thermal averages for every process ``$R$'', listed in Table~\ref{tab:energy_threshold_levels}, are obtained by weighted integration of the corresponding cross sections with Maxwellian velocity distributions at specific translational temperatures $7500 \, \mathrm{K} \le T \le 50000 \, \mathrm{K}$~\footnote{The rate coefficients originally listed in the Ames database spanned the temperature range $7500 - 25000\,\mathrm{K}$, but additional values were computed at higher temperatures for the present work} for the colliding species $\mathrm{N_2}\left(i\right)$, $\mathrm{N}$:
\begin{equation}
 \begin{split}
  \left\langle \sigma_R \cdot g \right\rangle & \left( T \right) = \frac{1}{\sqrt{\pi \, \mu_{\mathrm{N_2-N}}}} \left( \frac{2}{\mathrm{k_B} T} \right)^{3/2} \times \\
  & \times \int_{E_{R}}^{\infty} \sigma_R \left( E_t \right) \, \exp \left( - \frac{E_t}{\mathrm{k_B} T} \right) E_t \, \mathrm{d} E_t,
 \label{eq:maxwell_average}
 \end{split}
\end{equation}

where $g = \left| \boldsymbol{c}_\mathrm{N_2} - \boldsymbol{c}_\mathrm{N} \right|$ is the magnitude of the pre-collision relative velocity of the two collision partners, $\mu_{\mathrm{N_2-N}} = m_\mathrm{N_2} m_\mathrm{N} / \left( m_\mathrm{N_2} + m_\mathrm{N} \right)$ is their reduced mass and the relative translational energy is $E_t = \frac{1}{2} \mu_{\mathrm{N_2-N}} \, g^2$. Although theoretically the integration limits could range from zero to infinity, most of the cross sections are non-zero only within a certain energy range. The lower integration limit is given by $E_{R} \ge 0$, where the threshold value $E_{R}$ stands for the minimum amount of pre-collision translational energy necessary for the reaction products to be formed. According to the listing in Table~\ref{tab:energy_threshold_levels}, this quantity depends on the type of process and the specific rovibrational levels involved. Notice that we have assumed the energy threshold for dissociation from quasi-bound levels to be equal to zero (second row in Table~\ref{tab:energy_threshold_levels}). Strictly speaking, this ignores the small amount of energy initially needed to overcome the molecule's centrifugal barrier. However, this small error will not affect the evaluation of Eq.~(\ref{eq:maxwell_average}), since for such quasi-bound levels the dissociation cross section, and with it the integrand in Eq.~(\ref{eq:maxwell_average}), remains zero until reaching the true energy threshold.

The integral in Eq.~(\ref{eq:maxwell_average}) was computed numerically using the standard midpoint rule. The upper limit for integration was $E_{t, \mathrm{max}} = 50 \, \mathrm{eV}$, since cross sections in the database were only tabulated up to this value. However, given that practically all cross sections in the database were found to decrease to zero at energies above $30 \, \mathrm{eV}$, this finite upper bound did not constitute a problem.

\begin{table}
 \centering
 \caption{Energy threshold $E_{R}$ depending on type of process and level}
 \label{tab:energy_threshold_levels}
 
 \begin{tabular}{l | c | c}
   Process $R$ & $E_{R}$              & levels \\ \hline
  \multirow{2}{*}{Dissociation} & $2 E_\mathrm{N} - E_i$ & $i \in \mathcal{I}_\mathrm{B}$ \\
               & 0\*                      & $i \in \mathcal{I}_\mathrm{P}$ \\ \hline
    Excitation & $E_j - E_i$            & \multirow{3}{*}{$i,j \in \mathcal{I}_\mathrm{BP}$} \\
  Deexcitation & 0                      & \\
       Elastic & 0                      &
 \end{tabular}
\end{table}

Taking into account the energy threshold for each individual process from Table~\ref{tab:energy_threshold_levels}, their respective reaction rate coefficients correspond to: 
\begin{align}
 k_i^{Df} \left( T \right) & = \left\langle \sigma_i^D \cdot g \right\rangle \left( T \right), & i \in \mathcal{I}_\mathrm{BP}, \label{eq:dissociation_rate_coefficient_levels} \\
 k_{i \rightarrow j}^{E} \left( T \right) & = \left\langle \sigma_{i \rightarrow j}^E \cdot g \right\rangle \left( T \right), \quad E_j > E_i, & i,j \in \mathcal{I}_\mathrm{BP} \label{eq:excitation_rate_coefficient_levels} \\
 k_i^\mathrm{el} \left( T \right) & = \left\langle \sigma_{ii}^E \cdot g \right\rangle \left( T \right), & i \in \mathcal{I}_\mathrm{BP} \label{eq:elastic_rate_coefficient_levels}
\end{align}

Notice that Eq.~(\ref{eq:excitation_rate_coefficient_levels}) only accounts for excitation from lower rovibrational energy $E_i$ to higher energy $E_j$. The corresponding deexcitation rate coefficients are obtained directly as $k_{j \rightarrow i}^{E} \left( T \right) = g_i / g_j \exp \left[ \left(E_j - E_i \right) /\left( \mathrm{k_B} T \right) \right] k_{i \rightarrow j}^{E} \left( T \right)$, ensuring that detailed balance is satisfied for each pair of processes. The dissociation rate coefficients $k_i^{Df}$ and excitation rate coefficients $k_{i \rightarrow j}^{E}$ were first used by Panesi et al~\cite{panesi13a} to study the full reaction mechanism of the Ames N3 database by means of a set of master equations. A new addition in the present paper are the \emph{elastic} rate coefficients in the shape of Eq.~(\ref{eq:elastic_rate_coefficient_levels}). They do not appear in the chemical production terms of the master equations, because elastic collisions by definition have no effect on the production/depletion rates of individual rovibrational levels. The effect of elastic collisions only becomes relevant within the context of the hydrodynamic description, if transport phenomena such as viscous shear, heat conduction and multi-species diffusion are taken into account. On the other hand, within the kinetic description on which the DSMC method is based, elastic collisions must be accounted for explicitly as part of the physical modeling. In the DSMC collision routines the elastic cross sections need to be specified in addition to those for the inelastic and reactive processes. This issue has been addressed in different ways by other groups. Kim and Boyd~\cite{kim14a} opted for fitting the elastic cross section data from the Ames N3 database to an analytical form and derived 61 separate vibrational-state-specific total cross sections $\sigma_{v}^T \left( E_t \right)$ (i.e. sum of elastic, and all pseudo-reactions). Levin et al~\cite{parsons14a, li15a, zhu16a} instead derived a total cross section independent of the molecule's pre-collision internal state. They used the aggregate results of a large number of quasi-classical trajectories to determine the viscosity cross section $\sigma^\mu$. These data were then fitted an analytical expression and used as a good approximation of the total cross section $\sigma_{\mathrm{N_2-N}}^T \left( E_t \right)$. In both approaches the elastic collision rates are determined implicitly, as the difference between the total (i.e. sum of all processes) and the sum of the reactive collision rates. In our work, the elastic rate coefficients of Eq.~(\ref{eq:elastic_rate_coefficient_levels}) become relevant again, once we discuss the URVC bin model in Sec.~\ref{sec:urvc_binning}.

\begin{figure}
 \centering
 \includegraphics[width=1.0\columnwidth]{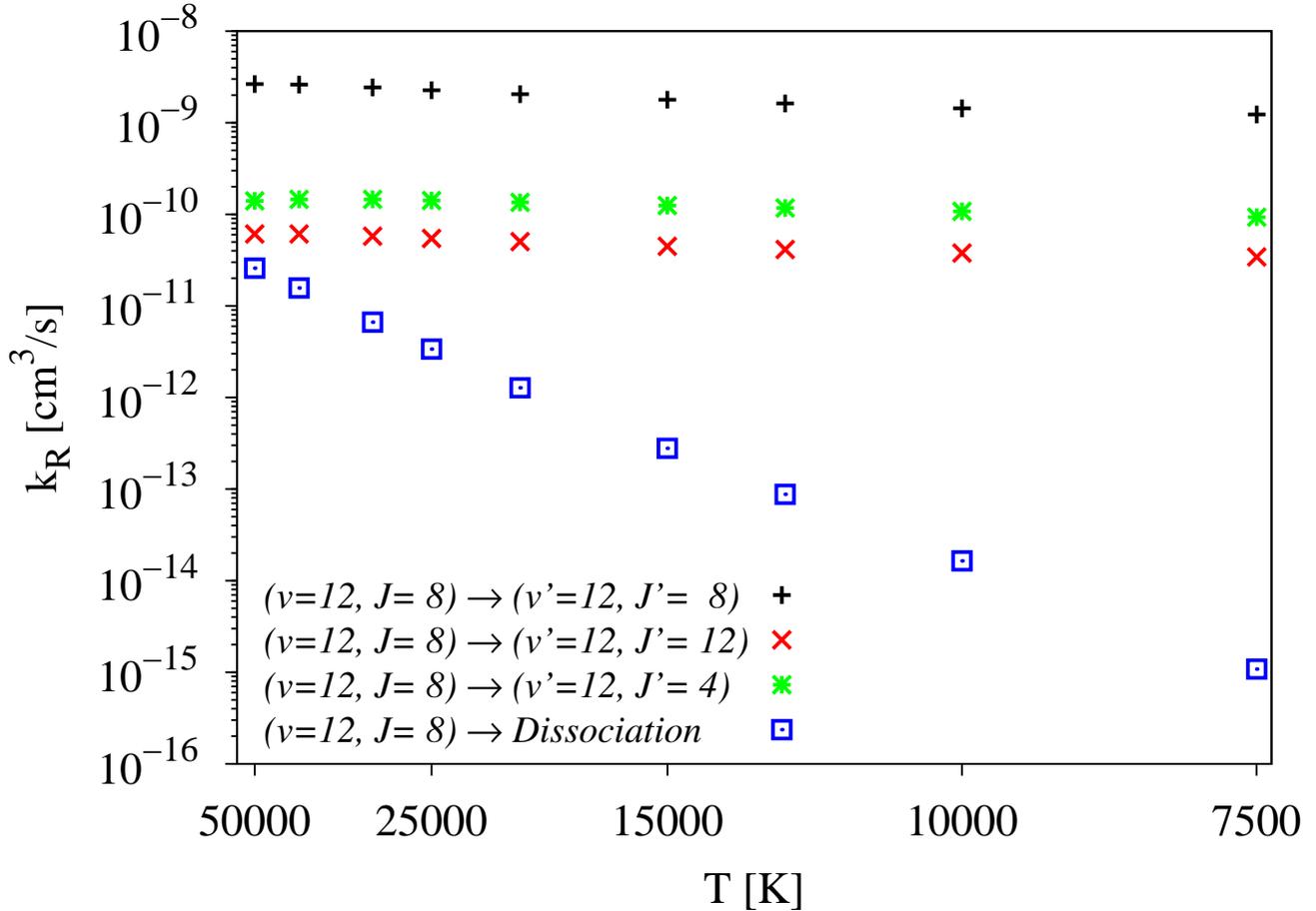}
 \caption{Numerically computed rate coefficients at selected temperatures for pre-collision level $(v = 12, J = 8)$. Black crosses: elastic rate coefficient; green asterisks: deexcitation to level $(v^\prime = 12, J^\prime = 4)$; red x's: excitation to level $(v^\prime = 12, J^\prime = 12)$; open blue squares: dissociation.}
 \label{fig:sample_rate_coefficients}
\end{figure}

The level-specific rate coefficients corresponding to the cross sections of Fig.~\ref{fig:sample_cross_sections} are plotted in Fig.~\ref{fig:sample_rate_coefficients}. Just like the cross sections they were derived from, these rate coefficients span many orders of magnitude. The largest one is again the elastic rate coefficient, while in this example both the excitation and deexcitation rate coefficients are roughly one order of magnitude smaller. The rate coefficients tend to increase with temperature. This is especially true for the dissociation rate coefficient, which increases 4 to 5 orders of magnitude between $7500$ and $50000 \, \mathrm{K}$. This behavior is a consequence of the relatively large energy threshold for dissociation from level ($v=12, J=18$). At temperatures below 10000~K approximately, the fraction of collisions in the gas with enough relative translational energy to surpass the dissociation barrier ($\Delta E_{\left(v=12, J=8\right)}^D = 6.5$ eV) is very small and the dissociation rate remains negligible. As the temperature increases, the maximum of the collision pair's joint velocity distribution shifts to higher energies, and with it the fraction of collisions possessing enough energy to dissociate increases as well. This in turn causes the dissociation rate coefficient to increase significantly at higher temperatures. One important fact is that, unlike the cross sections they are derived from, the rate coefficients exhibit a much smoother behavior over the temperature range considered. This is because the integration in Eq.~(\ref{eq:maxwell_average}) effectively causes the oscillations in the cross sections to be damped out and it is much easier to fit the rate coefficients to a simple analytical form, such as the well-known Arrhenius-type expressions.

We should note that experimental data for direct validation of the level-specific rate coefficients presented here does not exist. However, the global thermal rate coefficient for $\mathrm{N_2}$+$\mathrm{N}$-dissociation:
\begin{equation}
 k^{D} = \frac{1}{Q_\mathrm{N_2}^\mathrm{int}} \sum_{i\in\mathcal{I}_\mathrm{BP}} \left\lbrace g_i \, k_i^{Df} \exp \left( - \frac{E_i}{\mathrm{k_B} T} \right) \right\rbrace,
\end{equation}
which can be extracted from the Ames N3 database, has been found by Panesi et al~\cite{panesi13a} to agree well with Appleton's experimental results~\cite{appleton68a}.


\section{Uniform rovibrational collisional (URVC) energy bin model} \label{sec:urvc_binning}

In this section we propose several changes to the URVC bin model rate coefficients. The notation used here closely follows the one established by Magin et al.~\cite{magin12a}. However, for clarity some of the definitions will be repeated here. The bins are labeled with indices $k$ and $l$ for pre- and post-collision states respectively, and the set of all bin indices for molecular nitrogen will be labeled $\mathcal{K}_\mathrm{BP}$. Analogous to the convention introduced in Sec.~\ref{sec:rovibrational_levels}, this set is sub-divided into a set of bins containing only bound levels $\mathcal{K}_\mathrm{B}$ and another set of bins composed exclusively of pre-dissociated levels $\mathcal{K}_\mathrm{P}$. Thus, $\mathcal{K}_\mathrm{BP} = \mathcal{K}_\mathrm{B} \cup \mathcal{K}_\mathrm{P}$ and $\mathcal{K}_\mathrm{B} \cap \mathcal{K}_\mathrm{P} = \emptyset$. Quantities which refer to averages over a bin are designated with an over-bar. The binning procedure for the URVC model was explained by Magin et al~\cite{magin12a}, but the most important aspects will be recalled here, in addition to some proposed changes to their original formulation. 

Having sorted the rovibrational levels according to ascending energy, as shown in Fig.~\ref{fig:rovibrational_levels_energy}, a set of arbitrarily sized energy intervals $[\mathcal{E}_k, \mathcal{E}_{k+1})$ is defined for all bins $k \in \mathcal{K}_\mathrm{BP}$. In the original approach, the number and width of energy intervals could be freely chosen, except for the restriction that bound and quasi-bound levels were not to be mixed in the same bin. 

The number density of molecules belonging to bin $k$ is determined by summing over all levels involved: $\bar{n}_k = \sum_{i \in \mathcal{I}_k} \left\lbrace n_i \right\rbrace$, where $n_i$ is the number density of level $i$, and $\mathcal{I}_k$ is the set of indices pointing to all rovibrational levels $i$ contained within bin $k$: $\mathcal{I}_k = \left\lbrace i \in \mathcal{I}_\mathrm{BP} \; \text{such that} \; \left( \mathcal{E}_k \le E_i < \mathcal{E}_{k+1} \right) \right\rbrace$. The degeneracy of bin $k$, is given as the sum over all level degeneracies $g_i$ contained within the bin: $\bar{g}_k = \sum_{i \in \mathcal{I}_k} \left\lbrace g_i \right\rbrace$. A central assumption of the URVC bin model is that the individual level populations within each bin are fixed by the ratio of the level-to-bin degeneracies: $n_i / \bar{n}_k = g_i / \bar{g}_k$ and the bin energy $\bar{E}_k$ is defined as the weighted average over all level energies within the bin: $\bar{E}_k = 1 / \bar{g}_k \sum_{i \in \mathcal{I}_k} \left\lbrace g_i \, E_i \right\rbrace$.

\begin{figure}
 \centering
 \includegraphics[width=1.0\columnwidth]{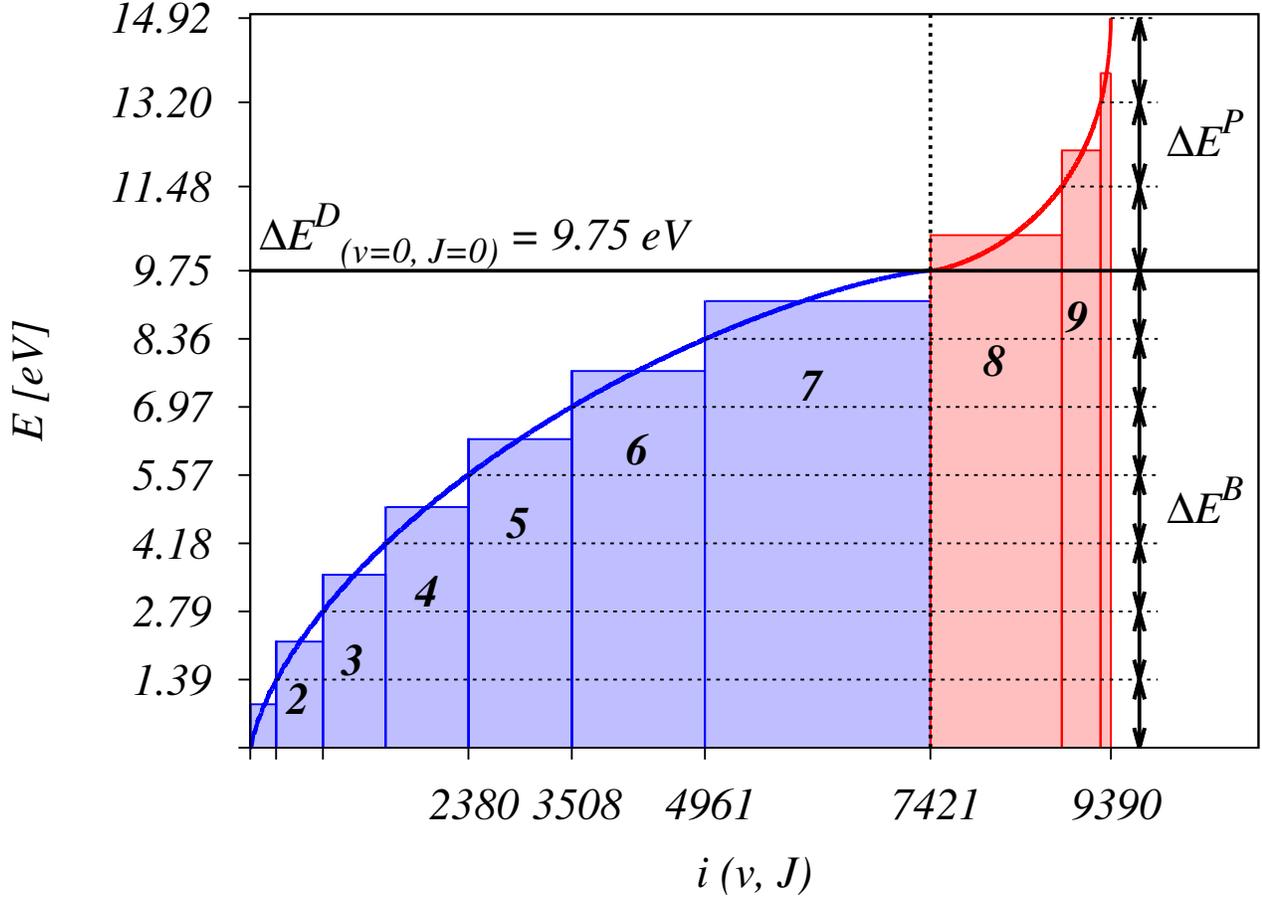}
 \caption{All 9390 rovibrational levels lumped into 10 energy bins. The first seven bins below the dissociation energy (blue boxes) are composed of bound levels. The upper three (red boxes) contain only quasi-bound ones. Equal-size energy intervals of width $\Delta E^B$ are used for all bins containing bound- and of width $\Delta E^P$ for all bins containing quasi-bound levels, respectively}
 \label{fig:7plus3_bins_equal_size_levels}
\end{figure}

In order to illustrate this procedure, take as an example a 10-bin system, where the first 7 bins are comprised only of bound levels and the remaining 3 bins contain only pre-dissociated levels. Assume now that the ``bound'' bins possess equally-spaced energy intervals $\Delta E^B = \mathcal{E}_{k+1} - \mathcal{E}_k, k \in \mathcal{K}_\mathrm{B}$ and the remaining pre-dissociated bins intervals $\Delta E^P = \mathcal{E}_{k+1} - \mathcal{E}_k, k \in \mathcal{K}_\mathrm{P}$. In this manner, each bin's boundaries are established and the sets $\mathcal{I}_k$ sorting levels into each bin follow automatically.
The properties of the resulting system are summarized in Table~\ref{tab:7plus3_bins_equal_size_numbers}, and a graphical representation is shown in Fig.~\ref{fig:7plus3_bins_equal_size_levels}. Here, each of the rectangles corresponds to a bin, with their ``height'' equal to $\bar{E}_k$ and their ``width'' given by the indices of the lowest- and highest-energy rovibrational levels belonging to that bin. The solid curve plotted on top of the bins represents the rovibrational level energies, repeated from Fig.~\ref{fig:rovibrational_levels_energy}.

\begin{table}
 \centering
 \caption{Average properties of the full set of levels lumped into 10 bins. The 7 lower bins are composed exclusively of bound levels, the upper 3 bins of quasi-bound levels}
 \label{tab:7plus3_bins_equal_size_numbers}

 \begin{tabular}{c | c | c | c | c}
  $\boldsymbol{k}$ & $\boldsymbol{i \in \mathcal{I}_k}$ & $\boldsymbol{\bar{g}_k}$ & $\boldsymbol{\mathcal{E}_k}, \boldsymbol{\mathcal{E}_{k+1}}$ \textbf{[eV]} & $\boldsymbol{\bar{E}_k}$ \textbf{[eV]} \\ \hline
  1 & \, \, 1 \ldots 282 &   76596 & 0.00 \ldots 1.39 & 0.89 \\
  2 & 283 \ldots 793 &  218817 & 1.39 \ldots 2.79 & 2.17 \\
  3 & \, 794 \ldots 1476 &  380712 & 2.79 \ldots 4.18 & 3.54 \\
  4 & 1477 \ldots 2380 &  585147 & 4.18 \ldots 5.57 & 4.92 \\
  5 & 2381 \ldots 3508 &  827016 & 5.57 \ldots 6.97 & 6.31 \\
  6 & 3509 \ldots 4960 & 1172946 & 6.97 \ldots 8.36 & 7.71 \\
  7 & 4961 \ldots 7421 & 1831185 & 8.36 \ldots 9.75 & 9.13 \\ \hline
  8 & 7422 \ldots 8854 & 1975419 & \, 9.75 \ldots 11.48 & 10.48 \\
  9 & 8855 \ldots 9277 &  869799 & 11.48 \ldots 13.20 & 12.21 \\
 10 & 9278 \ldots 9390 &  263841 & 13.20 \ldots 14.92 & 13.79 \\
 \end{tabular}
\end{table}

It should be noted that the definition of bins, which led to the system shown in Fig.~\ref{fig:7plus3_bins_equal_size_levels} is not unique, and other binning strategies are still being investigated (see Sec.~\ref{sec:introduction}).

\subsection{Updated definitions for bin-specific rate coefficients and equilibrium constants} \label{sec:urvc_binning_new_definitions}

As was the case in the original formulation~\cite{magin12a}, bin-averaged dissociation rate coefficients are obtained by grouping together all rate coefficients for the level-specific dissociation reactions of Eq.~(\ref{eq:dissociation_levels}):
\begin{equation}
 \bar{k}_k^{Df} = \frac{1}{\bar{g}_k} \sum_{i \in \mathcal{I}_k} g_i \, k_i^{Df} \qquad k \in \mathcal{K}_\mathrm{BP}, \label{eq:urvc_bin_N3_dissociation_rate_coefficient}
\end{equation}
where $\bar{k}_k^{Df}$ becomes the bin-averaged dissociation rate coefficient for the \emph{forward} process $\mathrm{N_2} \left( k \right) + \mathrm{N} \rightarrow 3 \, \mathrm{N}$. The bin-averaged recombination rate coefficient for the corresponding \emph{backward} reaction is determined by invoking detailed balance for the bin populations:
\begin{equation}
 \bar{k}_k^{Db} = \bar{k}_k^{Df} / \bar{K}_k^D, \qquad k \in \mathcal{K}_\mathrm{BP}, \label{eq:urvc_bin_N3_recombination_rate_coefficient}
\end{equation}

Contrary to the expression given in Appendix A of the original formulation~\cite{magin12a}, the equilibrium constant for this reaction is now defined as:
\begin{equation}
 \bar{K}_k^D = \frac{\left[ Q_\mathrm{N}^t \, g_\mathrm{N} \right]^2}{Q_\mathrm{N_2}^t \, \bar{g}_k} \, \exp \left( - \frac{2 E_\mathrm{N} - \bar{E}_k}{\mathrm{k_B} T} \right), \quad k \in \mathcal{K}_\mathrm{BP}, \label{eq:urvc_N3_bin_dissociation_equilibrium_constant}
\end{equation}
with the same expressions for the partition functions $Q_\mathrm{N}^t$ and $Q_\mathrm{N_2}^t$ as used in~\cite{magin12a}. The expression in Eq.~(\ref{eq:urvc_N3_bin_dissociation_equilibrium_constant}) ensures detailed balance for the dissociation- and recombination rates involving bin $k$ and atomic nitrogen. Notice that unlike in the original formulation~\cite{magin12a} any explicit reference to the rovibrational levels has disappeared from Eq.~(\ref{eq:urvc_N3_bin_dissociation_equilibrium_constant}) and that the dependence of the equilibrium constant on the original set of levels is now indirect, via the bin-averaged quantities $\bar{E}_k$ and $\bar{g}_k$.

In a similar fashion, weighted averaging is applied to the processes represented by Eq.~(\ref{eq:nonreactive_collisions_levels}), to produce bin-specific reaction rate coefficients for $\mathrm{N_2} \left( k \right) + \mathrm{N} \rightarrow \mathrm{N_2} \left( l \right) + \mathrm{N}$. Again, depending on the energies of the pre- and post-collision bins involved, three possibilities exist:

1) $\bar{E}_l > \bar{E}_k$: Excitation from bin $k$ to the higher-energy bin $l$. In this case, any pre-collision rovibrational levels $i \in \mathcal{I}_k$ will possess energies lower than those of the post-collision levels $j \in \mathcal{I}_l$:
\begin{equation}
 \bar{k}_{k \rightarrow l}^E = \frac{1}{\bar{g}_k} \sum_{i \in \mathcal{I}_k} g_i \sum_{j \in \mathcal{I}_l} k_{i \rightarrow j}^E, \qquad \bar{E}_k < \bar{E}_l. \label{eq:urvc_bin_N3_excitation_rate_coefficient}
\end{equation}

2) $\bar{E}_l < \bar{E}_k$: Deexcitation from bin $k$ to the lower-energy bin $l$. Any pre-collision rovibrational levels $i \in \mathcal{I}_k$ will possess energies higher than those of the post-collision levels $j \in \mathcal{I}_l$. For consistency, these deexcitation rate coefficients are computed indirectly, by invoking detailed balance for the bin populations:
\begin{equation}
 \bar{k}_{k \rightarrow l}^{E} = \bar{k}_{l \rightarrow k}^{E} / \bar{K}_{l \rightarrow k}^E, \qquad \bar{E}_k > \bar{E}_l. \label{eq:urvc_bin_N3_deexcitation_rate_coefficient}
\end{equation}

The coefficient $\bar{k}_{l \rightarrow k}^{E}$ on the right-hand-side of Eq.~(\ref{eq:urvc_bin_N3_deexcitation_rate_coefficient}) is the corresponding excitation rate coefficient for the transition $l \rightarrow k$, and the equilibrium constant relating the number densities of bins $k$ and $l$ is given by:
\begin{equation}
 \bar{K}_{l \rightarrow k}^E = \frac{\bar{g}_k}{\bar{g}_l} \exp \left( - \frac{\bar{E}_k - \bar{E}_l}{\mathrm{k_B} T} \right), \qquad k,l \in \mathcal{K}_\mathrm{BP}. \label{eq:urvc_bin_N3_bin_excitation_equilibrium_constant}
\end{equation}

The definition of this equilibrium constant differs from the original formulation given in Appendix A of Magin et al~\cite{magin12a} for the same reasons as the re-definition of $\bar{K}_k^D$ in  Eq.~(\ref{eq:urvc_N3_bin_dissociation_equilibrium_constant}). Instead of attempting to enforce detailed balance between all rovibrational levels via the equilibrium constant, which was not possible in the original formulation of the URVC bin model~\cite{magin12a}, Eq.~(\ref{eq:urvc_bin_N3_bin_excitation_equilibrium_constant}) is meant to only ensure that the bin populations $\bar{n}_k$ and $\bar{n}_l$ satisfy this condition. This change is necessary to ensure consistency with the detailed balance principle applied to the bin-specific excitation/deexcitation cross sections, expressed in Eq.~(\ref{eq:bin_excitation_detailed_balance_cross_sections}). 

3) $\bar{E}_l = \bar{E}_k$: Pre- and post-collision bins are the same. In this case some of the inelastic processes between different rovibrational levels, which happen to be lumped together into the same bin (i.e. those where $E_i \ne E_j$, with $i, j \in \mathcal{I}_k$), will be mixed together with the truly elastic collisions, i.e. those with $E_i = E_j$. When added together, all these processes will count toward the \emph{intra-bin} collision rate coefficient for bin $k$:
\begin{equation}
 \bar{k}_{k \rightarrow k}^\mathrm{coll} = \frac{1}{\bar{g}_k} \sum_{i \in \mathcal{I}_k} g_i \sum_{j \in \mathcal{I}_{k}} k_{i \rightarrow j}^E, \qquad k \in \mathcal{K}_\mathrm{BP}. \label{eq:urvc_bin_N3_elastic_rate_coefficient}
\end{equation}

This type of rate coefficient is a special case, since in our model intra-bin collisions do not directly influence the rates of dissociation and internal energy exchange. As a consequence, they do not appear in the chemical production terms $\bar{\omega}_\mathrm{N}$ and $\bar{\omega}_k$ given in Appendix A of the original formulation~\cite{magin12a}. They are nevertheless mentioned here, because intra-bin collisions are necessary to drive the local molecular velocities towards the Maxwell-Boltzmann equilibrium distribution and thus must be explicitly taken into account for a DSMC implementation.

As noted in Sec.~\ref{sec:level_specific_processes}, it should be recalled that the rovibrational-level-specific elastic cross sections $\sigma_{ii}^E$ extracted from the Ames database are not as reliable as those for the inelastic and reactive processes. Essentially, the elastic cross sections were only obtained as a by-product of the QCT-calculations. The original motivation for generating the N3-database was, after all, to determine the set of rovibrational state-to-state excitation- and state-specific dissociation rate coefficients for later use in classical CFD applications. Only impact parameters $0 \le b \le 4.2$\AA, were thus selected, with the certainty that the vast majority of the trajectories resulting in such inelastic/reactive processes would be found within this range. Then, from all the computed trajectories, only those for which the molecule exhibited no appreciable internal energy change, were classified as elastic. Increasing $b_\mathrm{max}$ beyond $4.2$\AA  would then only have added to the expense of the QCT calculations by producing mostly glancing trajectories, which no longer help to converge the inelastic cross sections. Bender et al~\cite{bender15a} used maximum impact parameters larger than $b_\mathrm{max} = 4.2$\AA~in their QCT calculations. However, they were studying $\mathrm{N_2}$-$\mathrm{N_2}$ interactions. In such a scenario, inelastic collisions are more likely to also occur at larger $b$-values, since both collision partners are rotating and vibrating diatoms, capable of disturbing each other's internal structure over greater distances. Despite this drawback, in this study we opted for still using the $k_{i}^\mathrm{el}, i \in \mathcal{I}_k$ from the database as part of the determination of $\bar{k}_{k\rightarrow k}^\mathrm{coll}$ in Eq.~(\ref{eq:urvc_bin_N3_elastic_rate_coefficient}). There, these elastic rate coefficients are lumped together with the additional inelastic ones $k_{i \rightarrow j}^E, i,j \in \mathcal{I}_k$, which also count toward the overall intra-bin collision rate coefficient of bin $k$.


\subsection{Fitting of binned rate coefficients to analytical form} \label{sec:urvc_arrhenius_fitting}

After computing the coarse-grain rate coefficients for the \emph{forward} excitation, intra-bin collision and dissociation reactions at several temperatures within the range $7500 \le T \le 50000 \, \mathrm{K}$, they were fit to a generic analytical expression of the form:
\begin{equation}
 k_R = A_R \, T^{b_R} \, \exp \left( - \frac{E_{R}}{\mathrm{k_B} \, T} \right). \label{eq:arrhenius_fit}
\end{equation}

For each reaction ``$R$'' the result is a pair of unique pre-exponential fitting parameters $A_R$ and $b_R$, in addition to the energy parameter $E_{R}$ in the exponential term. This energy parameter, although reaction-specific, is not obtained as a result of the fitting. In our approach, it is imposed beforehand. The value of $E_{R}$ marks the energy threshold below which the given reaction probability is assumed to be zero, and its value is calculated for a particular reaction according to the second column in Table~\ref{tab:energy_threshold_bins}.

In the original formulation of the URVC bin model~\cite{magin12a}, and in previous uses of the Ames database~\cite{panesi13a}, the same analytical form as in Eq.~(\ref{eq:arrhenius_fit}) had been used to fit the rate coefficient data. However, in those cases, $E_{R}$ would be treated as a third tuning parameter, determined by fitting the tabulated rate coefficients, and it would generally be different from the values given in Table~\ref{tab:energy_threshold_bins}. Switching to our two-parameter fitting becomes necessary in light of our goal to obtain cross sections in analytical form based on the fitted rate coefficients. This ``analytical inversion'' is discussed in Sec.~\ref{sec:cross_section_extraction}, where a key requirement is to enforce the precise value at which a given cross section switches from zero to non-zero. In this context, $E_{R}$ marks the threshold above which the total collision energy (i.e. translational plus internal) of the reactants starts to become greater than the minimum amount required to form the products. Given that in our coarse-grain model $\mathrm{N_2}$-molecules are now only allowed to possess a particular bin-averaged energy $\bar{E}_k$, these thresholds differ from those of the full set of rovibrational levels.

One should note at this point that Eq.~(\ref{eq:arrhenius_fit}) is only one of many functional forms one could use for fitting the rate coefficients. However, as will be made clear in Sec.~\ref{sec:cross_section_extraction}, a particular advantage of assuming Eq.~(\ref{eq:arrhenius_fit}) is that a simple, closed-form analytical expression for the corresponding cross section exists. This makes the task of obtaining cross sections from a large set of fitted rate coefficients straightforward. Although more complex forms for fitting rate coefficients exist, and in some cases may produce more accurate fits~\cite{zheng12a} than Eq.~(\ref{eq:arrhenius_fit}), they will in general not allow for such a closed-form analytical cross section. Since the main goal of this work is to generate a cross section set to be used in our DSMC solver, we will stick to Eq.~(\ref{eq:arrhenius_fit}) in the present paper.

In addition to $E_R$, Table~\ref{tab:energy_threshold_bins} lists the heat of reaction $\Delta E_R$ at $0 \, \mathrm{K}$ in the third column. Finally, the rightmost column specifies whether the previous values are meant for bins composed of bound, or quasi-bound levels. Notice that $E_R$ for dissociation from bin $k$ differs depending on whether the bin is composed of bound, or quasi-bound levels. In the first case, the reaction is endothermic, as indicated by $\Delta E_R$ being positive. Thus, the threshold $E_R$ has been set equal to the heat of reaction. By contrast, for bins composed of pre-dissociated levels $\Delta E_R < 0$, implying that the excess internal energy released during the reaction must be converted into relative translational energy of the newly-formed atoms. Therefore, in principle, a quasi-bound molecule already possesses enough energy to split into two atoms on its own. The additional translational energy supplied by the collision partner is only needed to initially overcome the molecule's centrifugal barrier. This fact was mentioned during the discussion of the rovibrational-level-specific processes of Table~\ref{tab:energy_threshold_levels}. One could argue then, that the threshold $E_R$ for dissociation from quasi-bound levels should be slightly positive, since the molecule must overcome the centrifugal barrier before being able to dissociate. However, given that the magnitude of this (positive) energy barrier is likely to be small compared to the (negative) heat of reaction, in our model we have set $E_R$ equal to zero for these processes.

For excitation from lower-energy bin $k$ to higher-energy bin $l$, the heat of reaction is positive, equal to the minimum amount of energy necessary to form the reaction products. Therefore, the energy threshold becomes $E_{k \rightarrow l}^E = \Delta E_{k \rightarrow l}^E = \bar{E}_l - \bar{E}_k$. In the opposite direction, deexcitation reactions are exothermic, and the energy threshold has been set equal to zero. Finally, no net internal energy changes occur during intra-bin collisions, and the corresponding energy threshold is automatically zero. All the \emph{forward}, i.e. the dissociation, excitation, or intra-bin collision rate coefficients are fitted to the analytical form of Eq.~(\ref{eq:arrhenius_fit}). The corresponding \emph{backward} deexcitation and recombination rate coefficients however, are computed indirectly by invoking detailed balance, as per Eqs.~(\ref{eq:urvc_bin_N3_recombination_rate_coefficient}) and (\ref{eq:urvc_bin_N3_deexcitation_rate_coefficient}) respectively. Given the definitions for the equilibrium constants of Eqs.~(\ref{eq:urvc_N3_bin_dissociation_equilibrium_constant}) and (\ref{eq:urvc_bin_N3_bin_excitation_equilibrium_constant}), and using the energy parameters listed in Table~\ref{tab:energy_threshold_bins}, these backward rate coefficients can also be written in simple analytical forms.

\begin{table}
 \centering
 \caption{Energy threshold $E_{R}$ and heat of reaction $\Delta E_R$ depending on type of process ``$R$'' and pre- and post collision bins $k$ and $l$ respectively}
 \label{tab:energy_threshold_bins}
 
 \begin{tabular}{l | c | c | c}
  Process ``$R$'' & $E_{R}$                                     & $\Delta E_R$                 & bins \\ \hline 
  Dissociation & $2 E_\mathrm{N} - \bar{E}_k$ & $2 E_\mathrm{N} - \bar{E}_k > 0$ & $k \in \mathcal{K}_\mathrm{B}$ \\
  from bin $k$ & 0                                             & $2 E_\mathrm{N} - \bar{E}_k < 0$ & $k \in \mathcal{K}_\mathrm{P}$ \\ \hline
  Recombination & 0 & $\bar{E}_k - 2 E_\mathrm{N} < 0$ & $k \in \mathcal{K}_\mathrm{B}$ \\
  to bin $k$ & $\bar{E}_k - 2 E_\mathrm{N}$ & $\bar{E}_k - 2 E_\mathrm{N} > 0$ & $k \in \mathcal{K}_\mathrm{P}$ \\ \hline
    Excitation $k \rightarrow l$ & $\bar{E}_l - \bar{E}_k$ & $\bar{E}_l - \bar{E}_k > 0$ & \multirow{3}{*}{$k,l \in \mathcal{K}_\mathrm{BP}$} \\
  Deexcitation $k \rightarrow l$ & 0 & $\bar{E}_l - \bar{E}_k < 0$ & \\
  Intra-bin coll. $k \rightarrow k$ & 0 & $\bar{E}_k - \bar{E}_k = 0$ &
 \end{tabular}
\end{table}

\begin{figure}
 \centering
 \includegraphics[width=1.0\columnwidth]{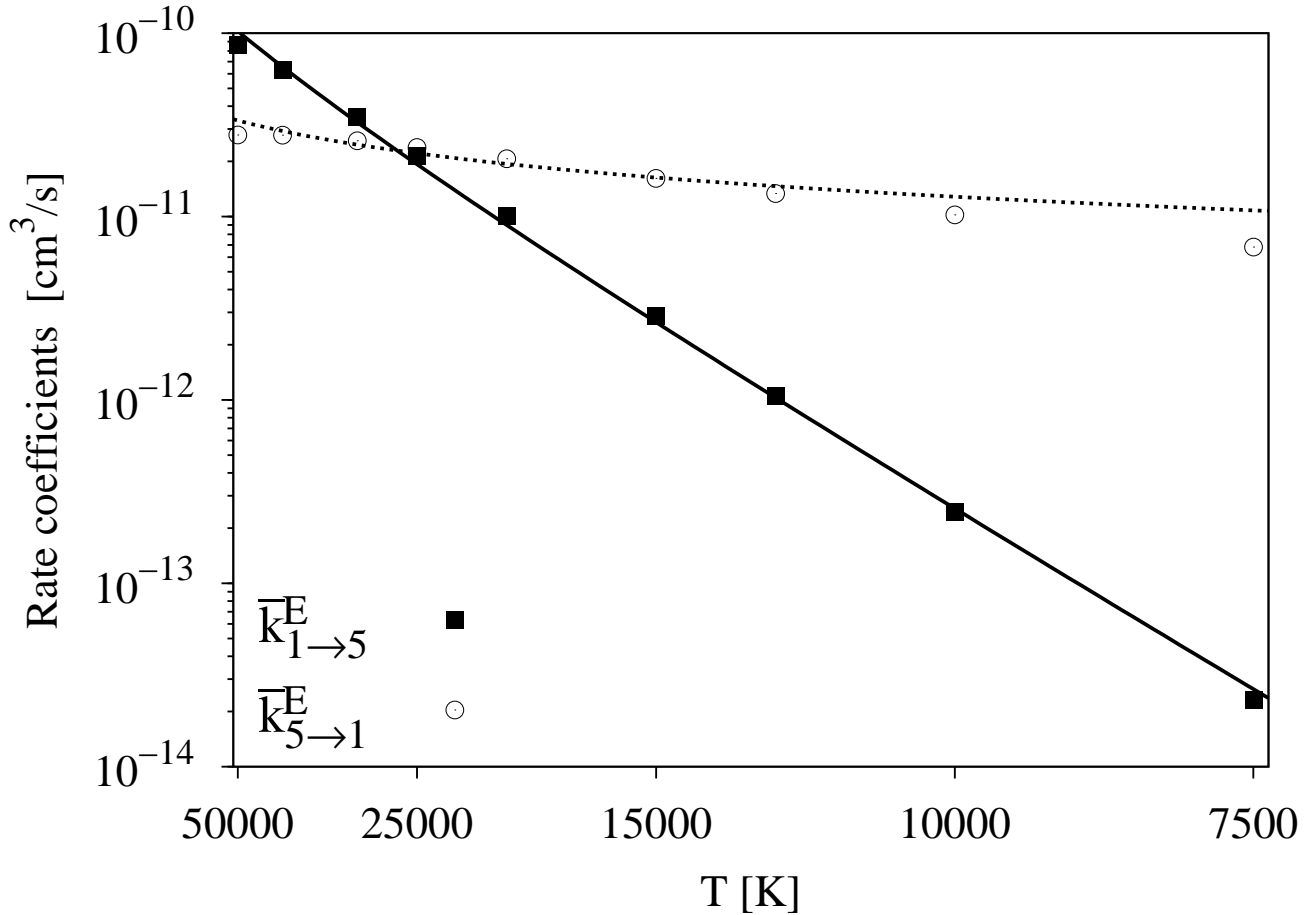}
 \caption{URVC excitation/deexcitation rate coefficients between bins 1 and 5 out of 10-bin system. Filled squares represent binned forward rate coefficients, while the open circles represent the corresponding backward rate coefficients for deexcitation. The solid line shows the fitted excitation rate coefficient, according to Eq.~(\ref{eq:arrhenius_fit}), with parameters $A_{1 \rightarrow 5}^E = 5.77 \times 10^{-13} \, \mathrm{cm^3/s}$, $b_{1 \rightarrow 5}^E = 0.595$. The dotted line for the corresponding deexcitation rate coefficient $\bar{k}_{5 \rightarrow 1}^E$ is obtained from Eq.~(\ref{eq:urvc_bin_N3_deexcitation_rate_coefficient_evaluated}).}
 \label{fig:rate_coefficients_bin_0001_0005}
\end{figure}

\begin{figure}
 \centering
 \includegraphics[width=1.0\columnwidth]{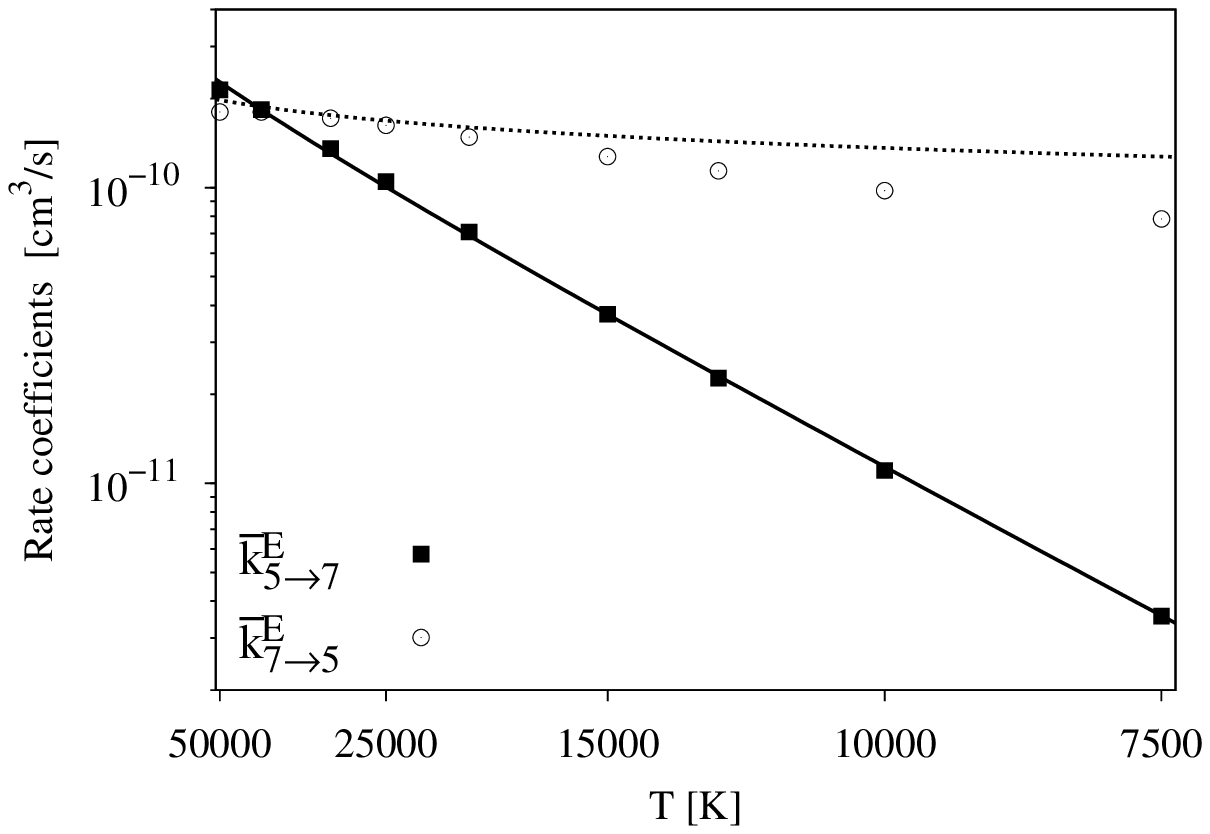}
 \caption{URVC excitation/deexcitation rate coefficients between bins 5 and 7 of 10-bin system. Filled squares represent binned forward rate coefficients, while the open circles represent the corresponding backward rate coefficients for deexcitation. The solid line shows the fitted excitation rate coefficient, according to Eq.~(\ref{eq:arrhenius_fit}), with parameters $A_{5 \rightarrow 7}^E = 3.56 \times 10^{-11} \, \mathrm{cm^3/s}$, $b_{5 \rightarrow 7}^E = 0.232$. The dotted line for the corresponding deexcitation rate coefficient $\bar{k}_{7 \rightarrow 5}^E$ is obtained from Eq.~(\ref{eq:urvc_bin_N3_deexcitation_rate_coefficient_evaluated}).}
 \label{fig:rate_coefficients_bin_0005_0007}
\end{figure}

\begin{figure}
 \centering
 \includegraphics[width=1.0\columnwidth]{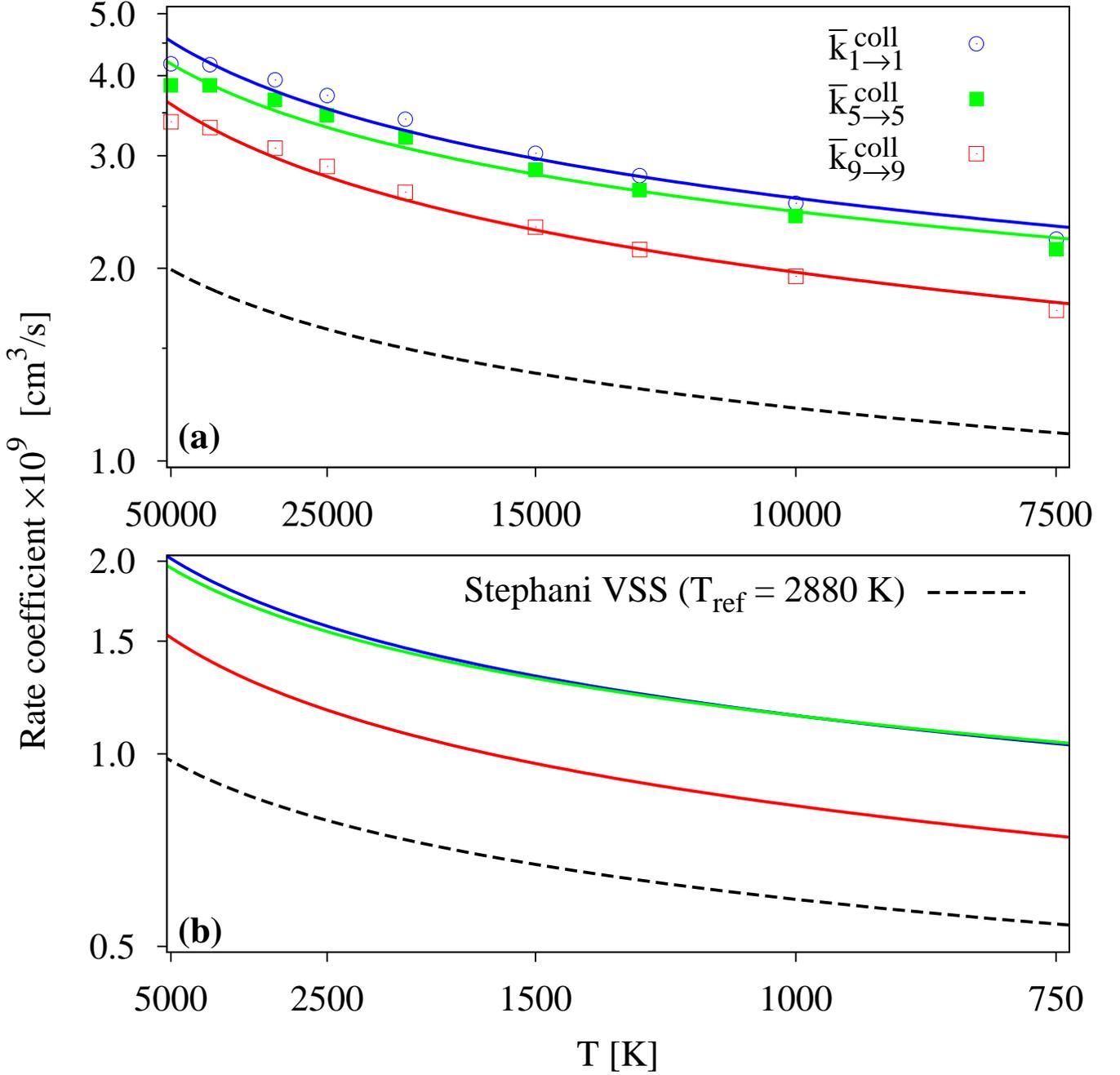}
 \caption{URVC intra-bin collision rate coefficients. In sub-figure (a) symbols represent binned rate coefficients for bin 1 (blue open circles), 5 (green filled squares) and 9 (red open squares) of 10-bin system. Solid lines of corresponding color represent the respective analytical fits according to Eq.~(\ref{eq:arrhenius_fit}). Sub-figure (b) shows extrapolation of fitted rate coefficients to lower temperatures. The dashed line in both plots represents $\left\langle \sigma_T^\mathrm{VHS} \cdot g \right\rangle \left( T \right)$ according to Eq.~(\ref{eq:vhs_collision_rate_coefficient}), with VHS parameters for $\mathrm{N_2}$-$\mathrm{N}$ taken from Stephani et al~\cite{stephani12a}.}
 \label{fig:rate_coefficient_bin_el}
\end{figure}

\begin{figure}
 \centering
 \includegraphics[width=1.0\columnwidth]{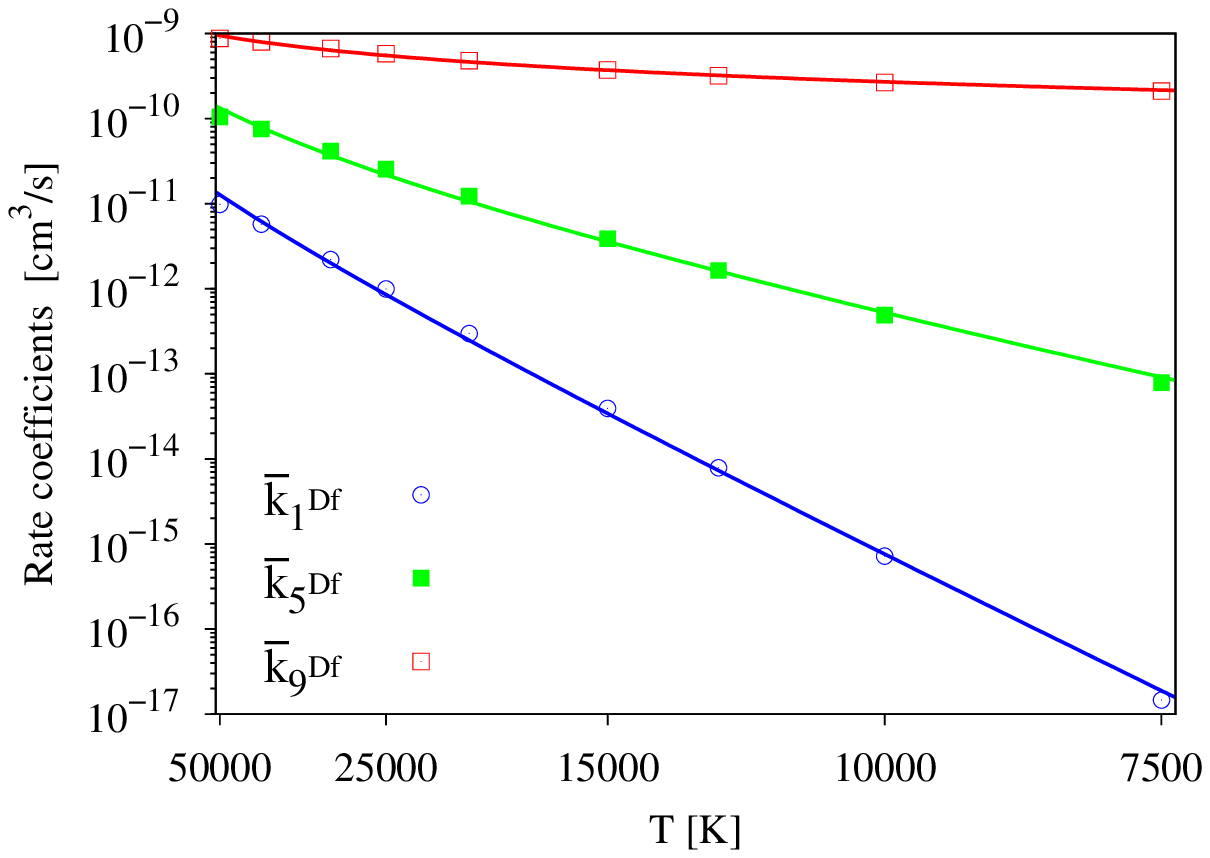}
 \caption{URVC dissociation rate coefficients for bins 1 (blue open circles), 5 (green filled squares) and 9 (red open squares) of 10. Symbols: binned rate coefficients, solid lines: analytical fits}
 \label{fig:rate_coefficients_bin_diss}
\end{figure}

A small sample of the rate coefficients belonging to the 10-bin system defined in Sec.~\ref{sec:urvc_binning} is shown in Figs.~\ref{fig:rate_coefficients_bin_0001_0005} to \ref{fig:rate_coefficients_bin_diss}. These serve to highlight the most important features observed for all processes (refer to App.~\ref{app:analytical_fit_parameters} for the complete list of fit parameters obtained for this 10-bin system). In all figures the rate coefficients tabulated at temperatures ranging from $7500$ to $50000 \, \mathrm{K}$ are represented by symbols, with the corresponding rate coefficients computed from our 2-parameter analytical fits shown as solid lines. The dotted lines in Figs.~\ref{fig:rate_coefficients_bin_0001_0005} and \ref{fig:rate_coefficients_bin_0005_0007} represent the rate coefficients for deexcitation in analytical form, which are obtained by combining Eqs.~(\ref{eq:arrhenius_fit}), (\ref{eq:urvc_bin_N3_deexcitation_rate_coefficient}) and (\ref{eq:urvc_bin_N3_bin_excitation_equilibrium_constant}) to yield:

\begin{equation}
 \bar{k}_{k \rightarrow l}^E = \frac{\bar{g}_l}{\bar{g}_k} \, A_{l \rightarrow k} \, T^{b_{l \rightarrow k}}, \qquad \bar{E}_k > \bar{E}_l, \quad k,l \in \mathcal{K}_\mathrm{BP}. \label{eq:urvc_bin_N3_deexcitation_rate_coefficient_evaluated}
\end{equation}

Notice that the exponential terms present in Eqs.~(\ref{eq:urvc_bin_N3_bin_excitation_equilibrium_constant}) and (\ref{eq:arrhenius_fit}) cancel out, resulting in the rather simple form of Eq.~(\ref{eq:urvc_bin_N3_deexcitation_rate_coefficient_evaluated}). It corresponds to a rate coefficient in analytical form with zero energy threshold, where $A_{k \rightarrow l} = A_{l \rightarrow k} \, \bar{g}_l / \bar{g}_k$ and $b_{k \rightarrow l} = b_{l \rightarrow k}$. As a consequence, the analytical deexcitation rate coefficients of Eq.~(\ref{eq:urvc_bin_N3_deexcitation_rate_coefficient_evaluated}) verify detailed balance for the bin populations, and are fully determined through the fitted parameters of the corresponding excitation rate coefficients. Focus first on Fig.~\ref{fig:rate_coefficients_bin_0001_0005}, where excitation and deexcitation rate coefficients between bins 1 and 5 are shown. The forward rate coefficient $\bar{k}_{1 \rightarrow 5}^E$ is shown as filled squares, while $\bar{k}_{5 \rightarrow 1}^E$ for the backward process is plotted as open circles. Notice that the excitation rate coefficient increases roughly four orders of magnitude with increasing temperature, whereas the deexcitation rate coefficient remains almost constant. This can be attributed to the large energy threshold $\bar{E}_{1 \rightarrow 5}^E= \bar{E}_5 - \bar{E}_1 = 5.42 \, \mathrm{eV}$ for excitation, as opposed to the reverse process, which requires zero energy threshold. For both $\bar{k}_{1 \rightarrow 5}^E$ and $\bar{k}_{5 \rightarrow 1}^E$ the analytical curves deviate from the tabulated values and slightly over-predict the excitation/deexcitation rates at the low- and high-temperature limits of the plot. Then, at temperatures between 15000 and roughly 30000~K, the analytical curves under-predict the tabulated values. Recall that the two analytical curves are linked via Eq.~(\ref{eq:urvc_bin_N3_deexcitation_rate_coefficient_evaluated}), and the tabulated values via Eq.~(\ref{eq:urvc_bin_N3_bin_excitation_equilibrium_constant}). Thus, the deexcitation rate coefficients mirror the behavior of the corresponding excitation rate coefficients, albeit at different magnitudes. However, since both sets of data are plotted on a logarithmic scale, the discrepancies between analytical and tabulated coefficients are more visible in the deexcitation rate coefficients, which only vary over one order of magnitude. The discrepancies seen here can attributed to the rather ``rigid'' 2-parameter analytical form, which with only two tunable values somewhat limits the ability of the analytical curves to follow the curvature of the tabulated values over a wider temperature range. However, as is seen in Sec.~\ref{sec:analytical_inversion} the 2-parameter analytical form is convenient for extracting analytical cross sections, which is why it is used it in this work.

In a similar fashion, Fig.~\ref{fig:rate_coefficients_bin_0005_0007} shows the pair of rate coefficients for excitation/deexcitation between bins 5 and 7 of the same 10-bin system. Notice that compared to the transitions $1 \rightleftharpoons 5$, both the forward and backward rate coefficients are much less sensitive to temperature. This can be attributed to the fact that with $\bar{E}_{5 \rightarrow 7}^E = \bar{E}_{7} - \bar{E}_{5} = 2.82 \, \mathrm{eV}$, the threshold for excitation is only half that of the previous case, which makes transitions between bins 5 and 7 much more probable, especially at lower temperatures. As was the case for the transition $1 \rightleftharpoons 5$, in Fig.~\ref{fig:rate_coefficients_bin_0005_0007} the tabulated (open circles) and fitted values (dotted line) of the deexcitation rate coefficient begin to noticeably deviate from one another as the temperature decreases. The same discussion as for Fig.~\ref{fig:rate_coefficients_bin_0001_0005} applies here, only that the differences are even more visible, given the narrower logarithmic range of the ordinate axis.

The intra-bin collision rate coefficients for bins 1, 5 and 9 of the 10-bin system are plotted in Fig.~\ref{fig:rate_coefficient_bin_el}. The fact that these are at least two orders of magnitude greater than the excitation rate coefficients previously shown is consistent with what was observed in Sec.~\ref{sec:level_specific_processes} for the rovibrational-level-specific cross sections and rate coefficients. Also recall from Eq.~(\ref{eq:urvc_bin_N3_elastic_rate_coefficient}), that $\bar{k}_{k \rightarrow k}^\mathrm{coll}$ is a sum of the purely elastic rate coefficients $k_{ii}^E$ and those inelastic rate coefficients $k_{i \rightarrow j}^E$, for which $i \ne j$, but both $i, j \in \mathcal{I}_k$. As before, the tabulated values for each of the bins are shown as symbols at temperatures between 7500 and 50000~K, with the solid lines of the same color representing the corresponding analytical fits. The fitted curves have been extrapolated down to the temperature range 750-5000~K in Fig.~\ref{fig:rate_coefficient_bin_el}~(b), since these types of collisions are expected to be the most prevalent in the low-temperature limit. In addition to the three intra-bin collision rate coefficients for bins 1, 5 and 9, the equivalent equilibrium collision rate coefficient based on a VHS/VSS total cross section~\cite{bird80a, koura91a} has been plotted in Fig.~\ref{fig:rate_coefficient_bin_el}.

\begin{equation}
 \left\langle \sigma_T^\mathrm{VHS} \cdot g \right\rangle = \frac{\pi d_\mathrm{ref}^2}{3/2 - \omega} \sqrt{\frac{8 \, \mathrm{k_B} T}{\pi \mu_\mathrm{N_2 - N}}} \left( \frac{T}{T_\mathrm{ref}} \right)^{1/2 - \omega} \label{eq:vhs_collision_rate_coefficient}
\end{equation}

The expression of Eq.~(\ref{eq:vhs_collision_rate_coefficient}) is obtained by inserting the functional form for a VHS/VSS total cross section of Eq.~(\ref{eq:vhs_cross_section}) into Eq.~(\ref{eq:maxwell_average}) and evaluating the resulting integral analytically. If we use parameters $d_\mathrm{ref} = 2.88$\AA, $\omega = 0.69$ and $T_\mathrm{ref} = 2880 \, \mathrm{K}$, for $\mathrm{N_2}$-$\mathrm{N}$ collisions found in literature (see Table III of Stephani et al~\cite{stephani12a}), the dashed line is obtained. These VSS data were originally optimized to match transport properties of a 5-species air mixture at temperatures between 1000-5000~K, roughly covering the temperature range in Fig.~\ref{fig:rate_coefficient_bin_el}~(b). We have included this curve and extrapolated it to the higher temperatures of Fig.~\ref{fig:rate_coefficient_bin_el}~(a) to make two points. 

First, over the whole temperature range our intra-bin collision rate coefficients (and the corresponding cross sections, see Fig.~\ref{fig:elastic_cross_section_comparison}) are about 1.5 to 2 times greater than the one predicted using the VSS data. This difference does not come as a surprise, given that the VSS parameters of Stephani et al~\cite{stephani12a} were derived from fitting transport collision integrals at temperatures only up to 10000~K~\cite{stallcop01a}. Such transport data generally do not take into account the fact that $\mathrm{N_2}$-molecules populating high-lying vibrational levels may present cross sections to the incoming $\mathrm{N}$-atom that are much greater than those of molecules in the vibrational ground state. Therefore, fitting these data results in total cross sections that are significantly smaller than the ones we show here. The second, related point is that in our case the intra-bin collision cross sections, as well as the corresponding overall intra-bin rate coefficients, are seen to not only be a function of the relative collision energy $E_t$ (what is usually assumed when using the standard VHS/VSS model used in DSMC), but to also depend on the pre-collision internal state(s) of the collision partners. In particular, the intra-bin collision rate coefficients of Fig.~\ref{fig:rate_coefficient_bin_el} are specific to each bin $k$.

Such a discrepancy was also observed by Kim and Boyd~\cite{kim14a}. In their analysis of the Ames N3 database they report vibrational-level specific total cross sections for $\mathrm{N_2}$-$\mathrm{N}$, which are several times greater than the ones obtained using standard VHS/VSS parameters. Similar conclusions were reached by Zhu et al~\cite{zhu16a}. As alluded to in Sec.~\ref{sec:level_specific_processes}, the internal energy state of the collision partners becomes important at the moment of determining the elastic- and total collision cross section $\sigma_T$ used in the DSMC implementation. Both references cited here deal with the problem in a different manner, but agree in the fact that a standard VHS cross section tuned to reproduce low-temperature transport coefficients will under-estimate the actual $\mathrm{N_2}$+$\mathrm{N}$ collision rate.


Finally, Fig.~\ref{fig:rate_coefficients_bin_diss} shows the comparison of rate coefficients for dissociation from bins 1 (open circles), 5 (filled squares) and 9 (open squares) respectively. As can be seen, these rate coefficient span many orders of magnitude over the temperature range considered. The trend is clear, with the strongest dependence on temperature for $\bar{k}_1^{Df}$, followed by $\bar{k}_5^{Df}$ and $\bar{k}_9^{Df}$. As will be further discussed in Sec.~\ref{sec:cross_sections_comparison}, where the corresponding cross sections are shown in Fig.~\ref{fig:cross_sections_comparison_diss}, this behavior can be mainly attributed to the significant differences in energy thresholds of the three processes.

\section{Extraction of URVC bin-specific cross sections from Ames database} \label{sec:cross_section_extraction}

We generate bin-averaged reaction cross sections for the processes in Table~\ref{tab:energy_threshold_bins}, in order to make use of the URVC bin model within the framework of a DSMC algorithm.

\subsection{Direct binning of level-specific cross sections} \label{sec:direct_binning}

The most straightforward way of obtaining such bin-average cross sections is to start from the formulas introduced in Sec.~\ref{sec:urvc_binning_new_definitions} for the rate coefficients, i.e. Eqs.~(\ref{eq:urvc_bin_N3_dissociation_rate_coefficient}), (\ref{eq:urvc_bin_N3_excitation_rate_coefficient}) and (\ref{eq:urvc_bin_N3_elastic_rate_coefficient}), and apply them instead to the original cross sections listed in the Ames database. For dissociation from bin $k$ this yields:
\begin{equation}
 \bar{\sigma}_k^{Df} = \frac{1}{\bar{g}_k} \sum_{i \in \mathcal{I}_k} g_i \, \sigma_i^{Df}, \qquad k \in \mathcal{K}_\mathrm{BP},
 \label{eq:direct_binning_diss}
\end{equation}
for excitation from bin $k$ to bin $l$:
\begin{equation}
 \bar{\sigma}_{k \rightarrow l}^E = \frac{1}{\bar{g}_k} \sum_{i \in \mathcal{I}_k} g_i \sum_{j \in \mathcal{I}_l} \sigma_{i \rightarrow j}^E, \quad \bar{E}_k < \bar{E}_l, \quad k, l \in \mathcal{K}_\mathrm{BP}, \label{eq:direct_binning_ex}
\end{equation}
and for intra-bin collisions we obtain:
\begin{equation}
 \bar{\sigma}_{k \rightarrow k}^\mathrm{coll} = \frac{1}{\bar{g}_k} \sum_{i \in \mathcal{I}_k} g_i \sum_{j \in \mathcal{I}_k} \sigma_{i \rightarrow j}^E \qquad k \in \mathcal{K}_\mathrm{BP}
 \label{eq:direct_binning_elastic}
\end{equation}

Although in principle this approach would require no additional manipulation of the original cross sections, several difficulties arise when applying these ``directly binned'' cross sections in a DSMC framework. First, notice that in determining every one of the bin-averaged cross sections of Eqs.~(\ref{eq:direct_binning_diss})-(\ref{eq:direct_binning_elastic}) one effectively lumps together level-specific cross sections spanning a wide range of different pre- and post-collision rovibrational levels. Take for example Eq.~(\ref{eq:direct_binning_diss}), for the bin-averaged dissociation cross section from bin $k$. Since every one of the level-specific dissociation cross sections in the set $\mathcal{I}_k$ possesses its own distinct energy threshold $E_{ i}^{Df} = 2 E_\mathrm{N} - E_i$, the resulting bin-average cross section $\bar{\sigma}_k^{Df}$ has its own energy threshold effectively ``smeared out'' over a finite range, instead of possessing a sharp threshold for becoming greater than zero at $\bar{E}_{k}^{Df} = 2 E_\mathrm{N} - \bar{E}_k$. The fact that the directly binned cross section could become non-zero for $E_t < \bar{E}_{k}^{Df}$ poses a problem when dealing with DSMC collision pairs $\mathrm{N_2}\left(k\right) + \mathrm{N}$ whose relative pre-collision translational energy lies slightly below this threshold. Such collision pairs would potentially be selected for reaction by the DSMC algorithm, even though they do not possess enough overall energy $E_t + \bar{E}_k$ to form the reaction products.

An additional inconvenience of this approach is that the directly binned cross sections would only be available in tabulated form at specific values of the collision energy $E_t$, as shown in Fig.~\ref{fig:sample_cross_sections}. During the course of a DSMC simulation the outcomes of many millions of individual collisions need to be determined, and the relative translational energy of each collision pair is a random value. Therefore, for most of the collisions it would be necessary to interpolate between tabulated values to obtain the precise value of the given cross section. These additional operations would add significantly to the overall run time of the simulation. Furthermore, the memory requirements for storing a large number of tabulated cross sections make this approach impractical for large-scale simulations. To alleviate these problems, an alternative method of obtaining the binned cross sections was investigated in Sec.~\ref{sec:analytical_inversion}.

\subsection{Analytical inversion of bin-specific rate coefficients} \label{sec:analytical_inversion}

It was proposed to use an analytical expression for the cross sections, based on the binned rate coefficients, by inverting Eq.~(\ref{eq:maxwell_average}). This approach may seem somewhat counter-intuitive, but is very similar to what has been common practice in conventional DSMC chemistry modeling (e.g. TCE method~\cite{bird78a, bird94a}). In this approach, the assumption is that all rate coefficients can be expressed using the analytical form of Eq.~(\ref{eq:arrhenius_fit}). As discussed in Sec.~\ref{sec:urvc_arrhenius_fitting}, the fitting parameters $A_R$ and $b_R$, as well as the energy threshold $E_{R}$ then take on particular values for the specific process in question. Next, an analytical expression with two adjustable parameters for the cross section of the following functional form is proposed~\footnote{It should be noted that analytical reaction cross section models have appeared in literature since the early 20th century~\cite{fowler49a}. For a recent compilation of analytical cross section models relevant for high-temperature processes in reentry flows, see Chernyi et al~\cite{chernyi02a}.}:
\begin{equation}
 \sigma_R \left( E_t \right) =
 \begin{cases}
  \displaystyle \frac{C_{R} \left( E_t - E_{R} \right)^{\eta_{R}}}{E_t} \enspace \text{if} \enspace E_t > E_{R}, \\
  \\
  0 \qquad \qquad \qquad \qquad \text{otherwise}.
 \end{cases}
 \label{eq:cross_section}
\end{equation}

Substituting Eq.~(\ref{eq:cross_section}) into the general expression of Eq.~(\ref{eq:maxwell_average}) turns out to exactly yield the functional dependence on temperature as the Arrhenius-type rate coefficient given by Eq.~(\ref{eq:arrhenius_fit}). After comparing terms on both sides of the equation, this automatically determines the two parameters:
\begin{align}
 \nonumber C_{R} & = \frac{A_R}{2 \cdot \Gamma \left( 3/2 + b_R \right)} \sqrt{\frac{\pi \, \mu_{\mathrm{N_2,N}}}{2}} \, \mathrm{k_B}^{-b_R}, \\
 \label{eq:cross_section_parameters} \\
 \nonumber\eta_{R} & = b_R + 1/2.
\end{align}

Thus, all cross sections obtained in this manner effectively share the same functional form and the precise shape of each individual cross section is only controlled by the fitting parameters $A_R$, $b_R$, as well as the reaction-specific energy threshold $E_{R}$, according to Table~\ref{tab:energy_threshold_bins}. Given binned rate coefficients in the form of Eq.~(\ref{eq:arrhenius_fit}), the fitting parameters are obtained for all \emph{forward} processes, i.e. excitation and dissociation reactions, as well as for the intra-bin collisions. The cross sections for all \emph{backward} processes, i.e. the deexcitation reactions, are obtained after an additional step. In analogy to the level-specific excitation/deexcitation cross sections discussed in Sec.~\ref{sec:level_specific_processes}, the set of bin-specific cross sections must verify micro-reversibility relations. 
Re-writing Eq.~(\ref{eq:detailed_balance_levels}) for the deexcitation cross section from higher-energy bin $k$ to lower-energy bin $l$ yields:
\begin{equation}
  \bar{\sigma}_{k \rightarrow l}^E \left( E_t \right) = \frac{\bar{g}_l \, E_t^\prime}{\bar{g}_k \, E_t} \, \bar{\sigma}_{l \rightarrow k}^E \left( E_t^\prime \right), \quad \bar{E}_k > \bar{E}_l \in \mathcal{K}_\mathrm{BP}, \label{eq:bin_excitation_detailed_balance_cross_sections}
\end{equation}
where the translational energy after deexcitation is equal to $E_t^\prime = E_t + \bar{E}_k - \bar{E}_l$, while $\bar{\sigma}_{l \rightarrow k}^E \left( E_t^\prime \right)$ is the cross section for the corresponding excitation reaction from lower-energy bin $l$ to higher-energy bin $k$, evaluated at $E_t^\prime$. Notice that the difference $\Delta E_{l \rightarrow k}^E = \bar{E}_k - \bar{E}_l$ is equal to the energy threshold $E_{ lk}^E$ required by this excitation reaction. Substituting all terms into Eq.~(\ref{eq:bin_excitation_detailed_balance_cross_sections}), yields the analytical form shared by all deexcitation cross sections in our model:
\begin{equation}
 \bar{\sigma}_{k \rightarrow l}^E \left( E_t \right) = \frac{\bar{g}_l}{\bar{g}_k} \, C_{l \rightarrow k}^E \, E_t^{\eta_{l \rightarrow k}^E - 1}, \quad \bar{E}_k > \bar{E}_l \in \mathcal{K}_\mathrm{BP}. \label{eq:bin_deexcitation_cross_sections_substituted}
\end{equation}

This functional form is consistent with Table~\ref{tab:energy_threshold_bins}, in the sense that the energy threshold for all deexcitation reactions is automatically zero. In our DSMC implementation of the URVC bin model~\cite{torres15a}, we have used Eq.~(\ref{eq:bin_excitation_detailed_balance_cross_sections}) to compute the bin-specific deexcitation cross sections consistent with the corresponding ones for excitation. In a similar fashion, Zhu et al~\cite{zhu16a} follow this approach to ensure micro-reversibility in their DSMC implementation of the 2D-bin model.

\begin{figure}
 \centering
 \includegraphics[width=1.0\columnwidth]{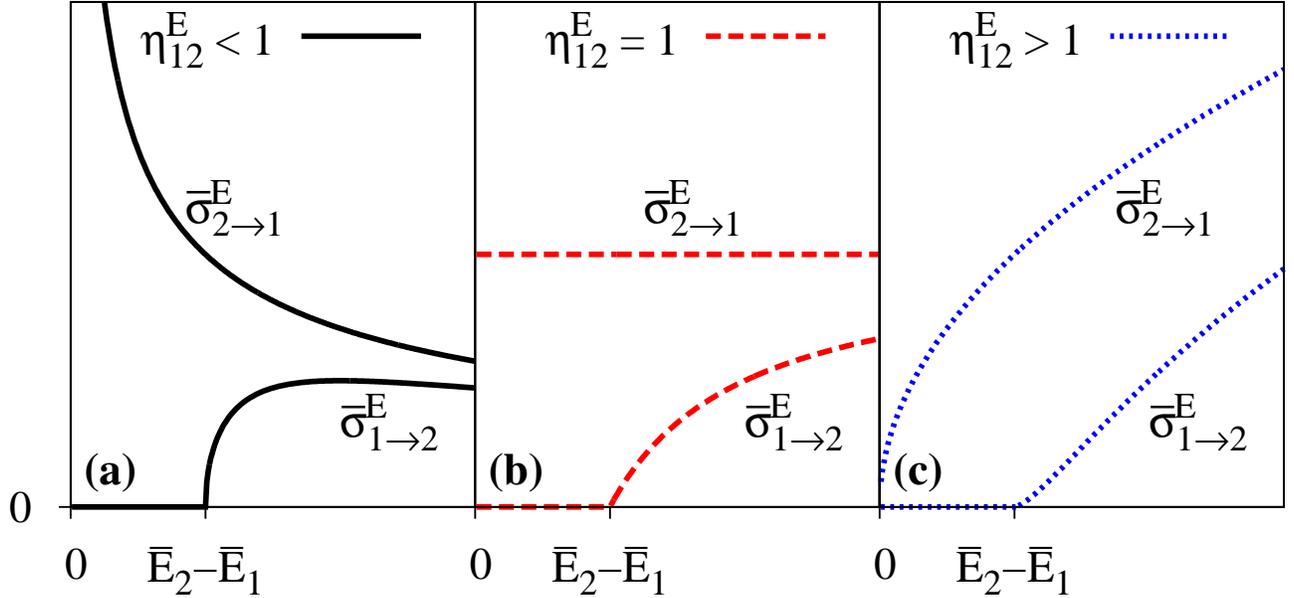}
 \caption{Characteristic shapes of corresponding analytical excitation/deexcitation cross sections between bins lower-energy bin 1 and higher-energy bin 2 depending on the value of $\eta_{1 \rightarrow 2}^E$}
 \label{fig:example_cross_sections}
\end{figure}

The shapes of both the excitation and deexcitation cross sections are determined by the values of the parameters $C_{l \rightarrow k}^E$ and $\eta_{l \rightarrow k}^E$. While the coefficient $C_{l \rightarrow k}^E$ controls the magnitude of both cross sections, the exponent $\eta_{l \rightarrow k}^E$ is responsible for the curvature and asymptotic behavior at very small- and at very large translational energies. This is illustrated in Fig.~\ref{fig:example_cross_sections}, with the example of excitation/deexcitation between lower-energy bin ``1'' and higher-energy bin ``2''. Depending on the value of $\eta_{1 \rightarrow 2}^E$, three types of behavior are possible. On the left of Fig.~\ref{fig:example_cross_sections}, if $\eta_{1 \rightarrow 2}^E < 1$, the deexcitation cross section $\bar{\sigma}_{2 \rightarrow 1}^E$ diverges as $E_t \rightarrow 0$. Simultaneously, the excitation cross section $\bar{\sigma}_{1 \rightarrow 2}^E$ exhibits a local maximum at $E_t^0 = \left( \bar{E}_2 - \bar{E}_1 \right) / \left( 1 -\eta_{1 \rightarrow 2}^E \right)$. Both cross sections slowly tend toward zero in the limit $E_t \rightarrow \infty$. The plot in the middle shows the behavior for $\eta_{1 \rightarrow 2}^E = 1$. The deexcitation cross section remains constant at $\bar{\sigma}_{2 \rightarrow 1}^E = \bar{g}_1 / \bar{g}_2 C_{1 \rightarrow 2}$, whereas the excitation cross section tends toward $C_{1 \rightarrow 2}$ as $E_t \rightarrow \infty$. This case is unlikely to occur when processing the binned rate coefficients, because it would only occur for the value of $b_{1 \rightarrow 2}^E$ being exactly equal to $1/2$. Finally, in the plot on the right, we sketch the situation for $\eta_{1 \rightarrow 2}^E > 1$. In this case, the deexcitation cross section starts at zero at $E_t = 0$ and then grows without bounds for $E_t \rightarrow \infty$, similar to the deexcitation cross section. Notice that in all three cases the cross section for excitation $\bar{\sigma}_{1 \rightarrow 2}^E$ remains zero until the threshold $E_{1 \rightarrow 2}^E = \bar{E}_2 - \bar{E}_1$ has been reached. As is shown in Sec.~\ref{sec:cross_sections_comparison}, in practice both $\eta_R < 1$ and $\eta_R > 1$ occur when inverting the binned rate coefficients.

Before moving on to a comparison of cross sections obtained through the direct binning and analytical inversion approaches, it is worth discussing two additional points. First, recall that in Sec.~\ref{sec:urvc_binning_new_definitions} the principle of detailed balance for excitation/deexcitation reactions was expressed by Eqs.~(\ref{eq:urvc_bin_N3_deexcitation_rate_coefficient}) and (\ref{eq:urvc_bin_N3_bin_excitation_equilibrium_constant}), which constitute its equivalent form at the hydrodynamic scale. By inserting Eq.~(\ref{eq:bin_excitation_detailed_balance_cross_sections}) into Eq.~(\ref{eq:maxwell_average}) and evaluating the integral, one can show that Eq.~(\ref{eq:urvc_bin_N3_deexcitation_rate_coefficient}) is indeed obtained, as long as the equilibrium constant takes on the form of Eq.~(\ref{eq:urvc_bin_N3_bin_excitation_equilibrium_constant}). The added value of Eq.~(\ref{eq:bin_excitation_detailed_balance_cross_sections}) is that it is valid at the kinetic scale, and thus needs to be respected when generating the reaction cross sections for the DSMC method. Second, it should be noted that several alternative methods for extracting cross sections from temperature-dependent rate coefficients exist. They are listed here for comparison, but were not used in the present work:

1) The inverse Laplace transform used by Kustova et al~\cite{baykov16a} is another method for obtaining cross sections in analytical form. It is based on the realization that the rate coefficient obtained via Eq.~(\ref{eq:maxwell_average}) can be interpreted as a Laplace transform of $\sigma_R$. It is similar to the inversion method presented here, and can be viewed as a generalization of our approach. Therefore, it also requires one to propose a specific functional form for the cross sections, which will allow them to match the temperature-dependence of the rate coefficients. However, care must be taken in selecting this functional form. Otherwise the inverse Laplace transform may yield nonphysical results, where the cross sections could become negative~\cite{kustova12a}.

2) Tikhonov regularization~\cite{tikhonov77a}, also known as the Phillips–Twomey method, is a numerical technique for the regularization of ill-posed problems. It was used by Bondar and Ivanov~\cite{bondar07a} to determine the reaction probabilities for DSMC, which would match the behavior of a two-temperature dissociation rate coefficient. This approach, although more flexible than the inverse Laplace transform, relies on an iterative numerical method for finding a plausible shape for the cross section. Since the problem is ill-posed, there is no prior guarantee that the obtained cross sections will be ``well-behaved'' and considerable user input is necessary in each individual case. Since the detailed-chemistry mechanisms considered in the present work involve hundreds, or even thousands of individual cross sections this approach was deemed impractical.
 
3) Minelli et al~\cite{minelli11a} used the non-linear constrained optimization method of Nelder and Mead~\cite{nelder65a}, aka. Downhill-simplex method. As a numerical technique, this approach shares many advantages and drawbacks with Tikhonov regularization. Similarly, finding each individual cross section requires considerable user surveillance and is therefore not well-suited to the present application.


\subsection{Comparison of directly binned and analytically inverted cross sections} \label{sec:cross_sections_comparison}

\begin{figure}
 \centering
 \includegraphics[width=\columnwidth]{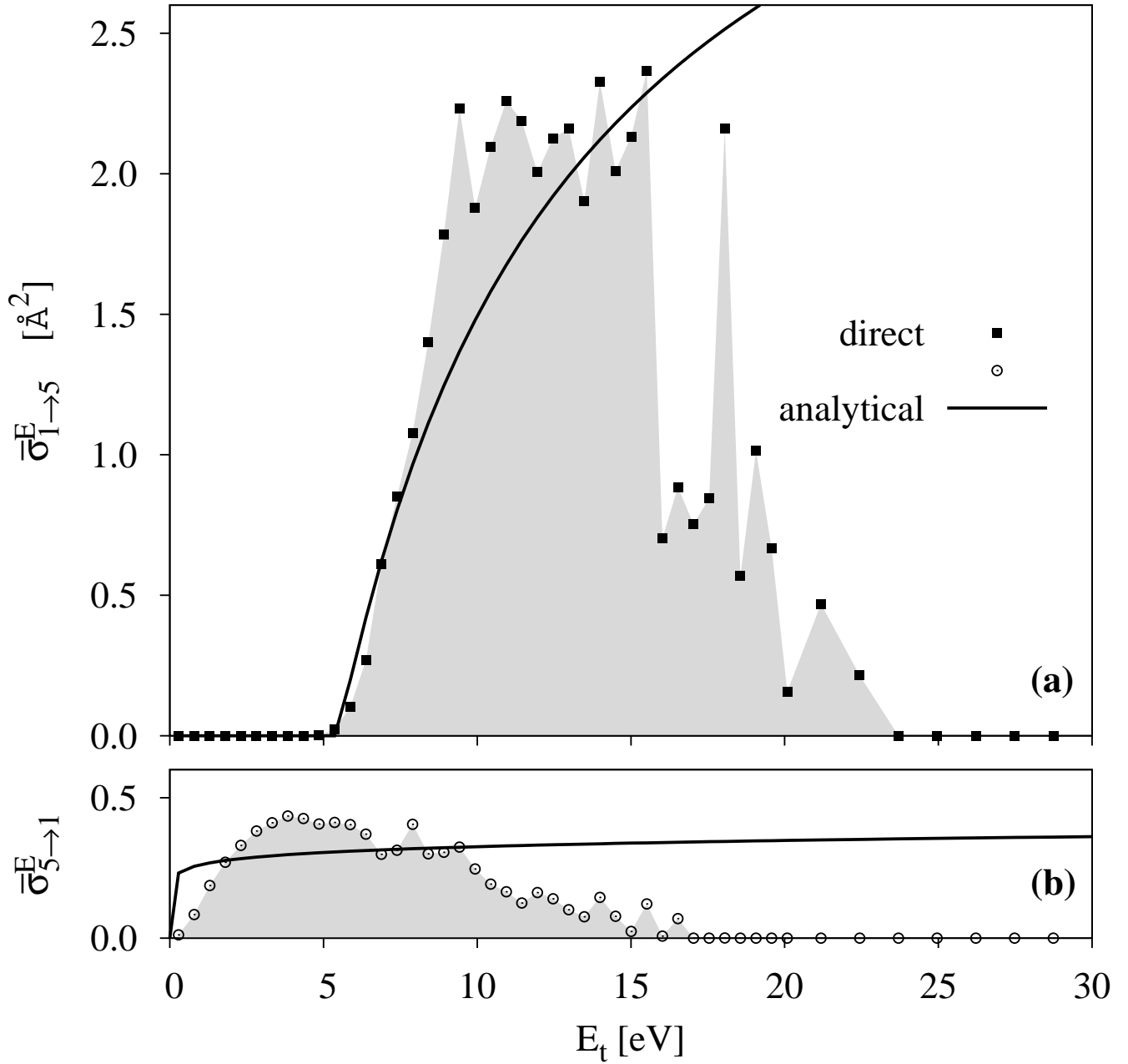}
 \caption{URVC-binned excitation/deexcitation cross sections between bins 1 and 5 of a 10. Symbols on shaded area: directly binned QCT cross section; solid lines: cross section obtained by analytical inversion.}
 \label{fig:cross_sections_comparison_15}
\end{figure}

\begin{figure}
 \centering
 \includegraphics[width=\columnwidth]{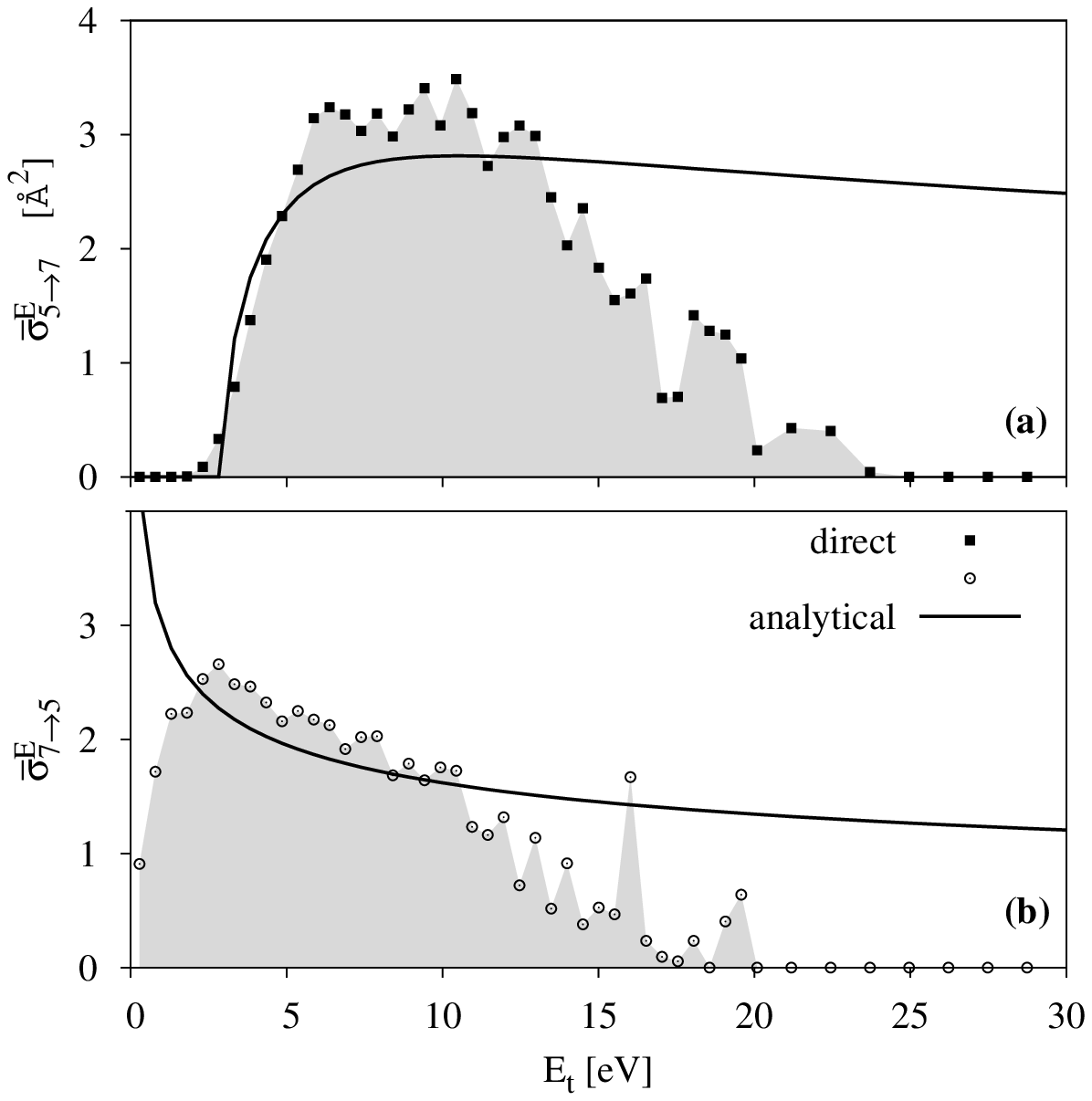} 
 \caption{URVC-binned excitation/deexcitation cross sections between bins 5 and 7 of 10. Symbols on shaded area: directly binned QCT cross section; solid lines: cross section obtained by analytical inversion.}
 \label{fig:cross_sections_comparison_57}
\end{figure}

\begin{figure}
 \centering
 \includegraphics[width=0.95\columnwidth]{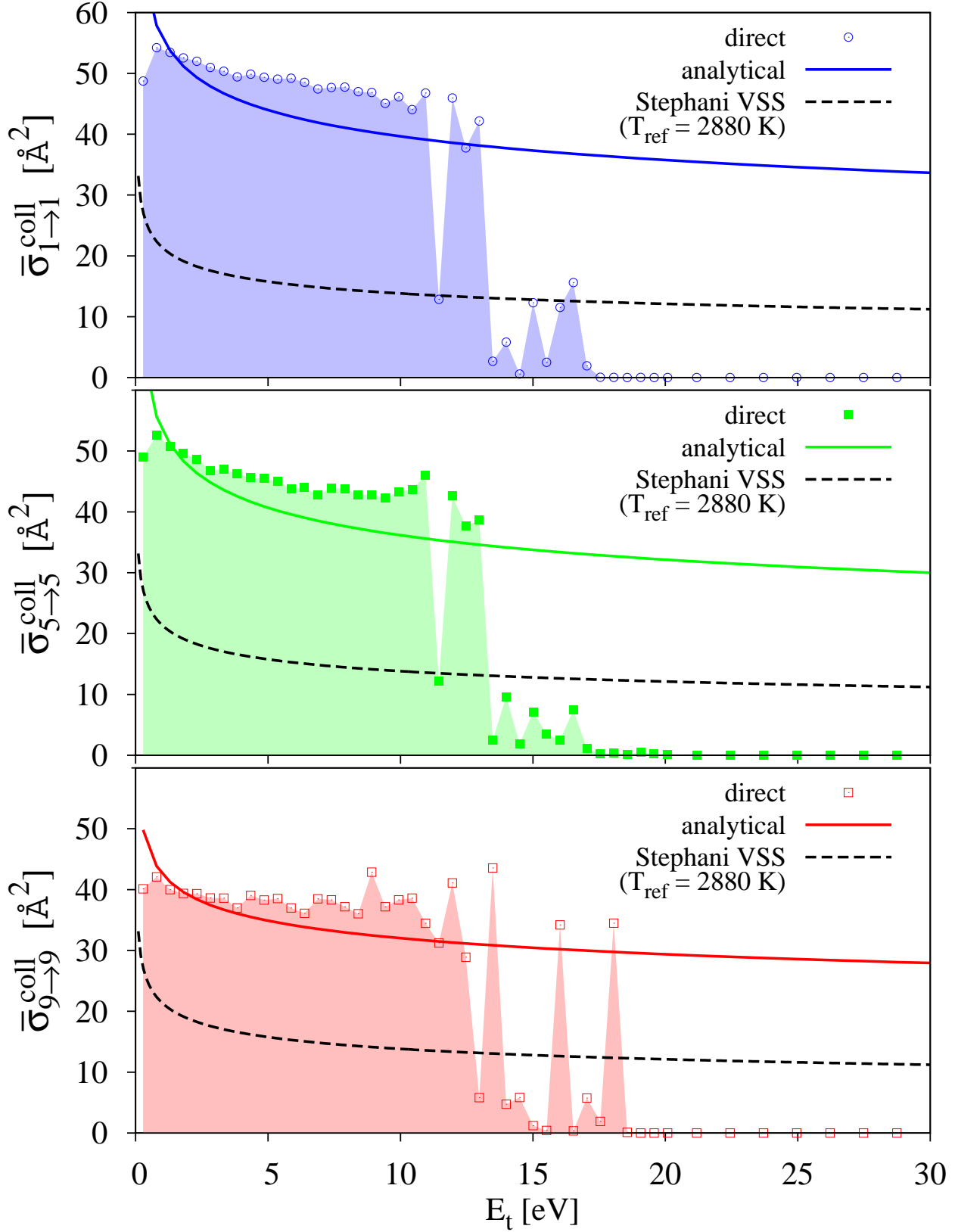} 
 \begin{singlespace*}
 \caption{Intra-bin collision cross sections for bins 1, 5 and 9 out of 10-bin system. Blue open circles on shaded area in sub-figure (a), green filled squares on shaded area in sub-figure (b) and red open squares on shaded area in sub-figure (c) represent the directly binned QCT cross sections $\bar{\sigma}_{1 \rightarrow 1}^\mathrm{coll}$, $\bar{\sigma}_{5 \rightarrow 5}^\mathrm{coll}$ and $\bar{\sigma}_{9 \rightarrow 9}^\mathrm{coll}$ respectively. In all three sub-figures the solid line shows the corresponding cross section obtained analytically through the combination of Eqs.~(\ref{eq:cross_section}) and (\ref{eq:cross_section_parameters}). Also, in all three sub-figures, the dashed lines represent $\sigma_\mathrm{VHS}^T \left( E_t \right)$ according to Eq.~(\ref{eq:vhs_cross_section}), with VHS parameters for $\mathrm{N_2}$-$\mathrm{N}$ taken from Stephani et al~\cite{stephani12a}.}
 \end{singlespace*}
 \label{fig:elastic_cross_section_comparison}
\end{figure}

\begin{figure}
 \centering
 \includegraphics[width=\columnwidth]{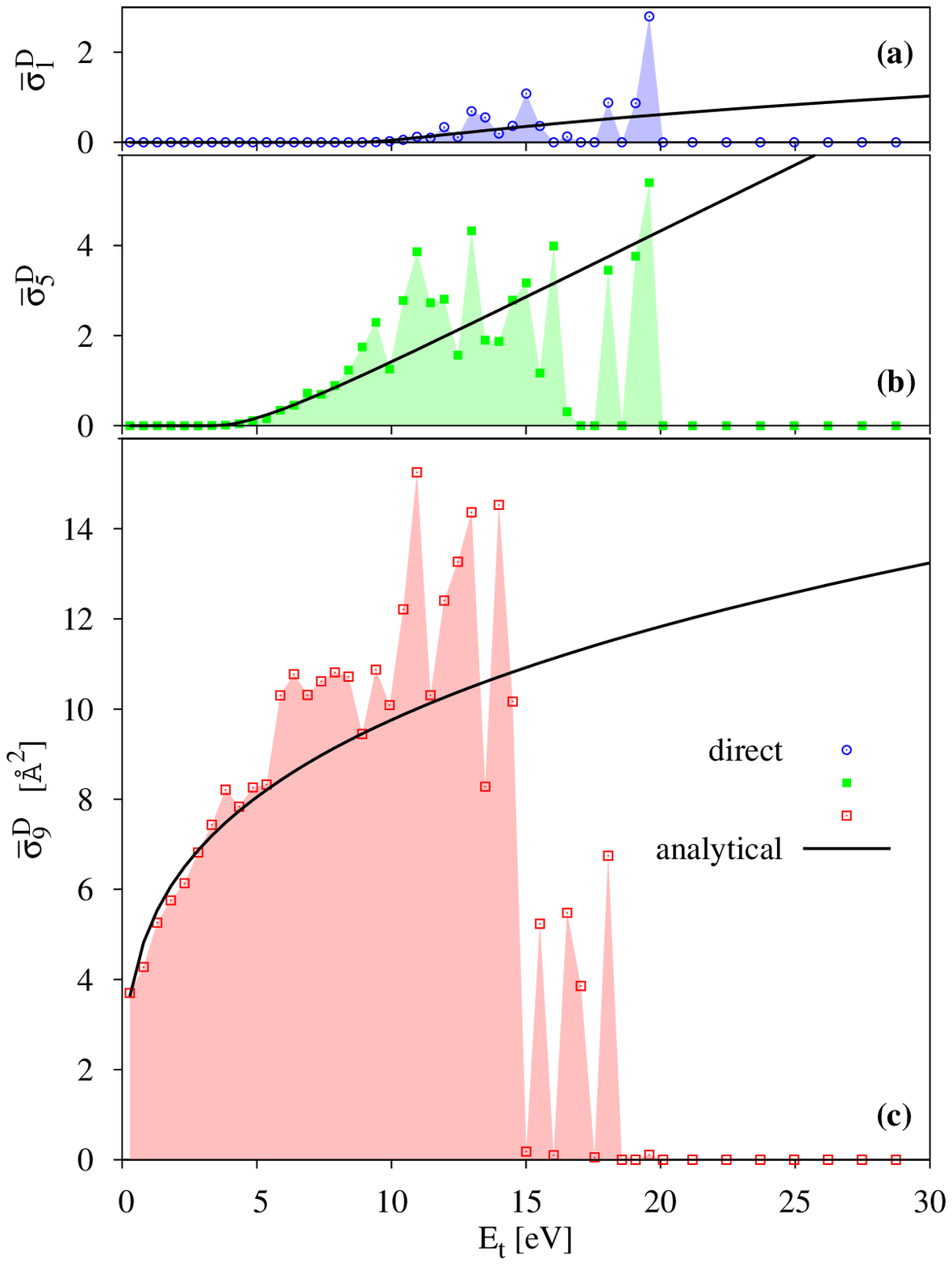}
 \caption{Dissociation cross sections for bins 1, 5 and 9 out of 10-bin system. Squares on shaded area: directly binned QCT cross section; solid line: cross section obtained by analytical inversion.}
 \label{fig:cross_sections_comparison_diss}
\end{figure}

Several example cross sections, obtained using the analytical inversion technique described in Sec.~\ref{sec:analytical_inversion}, are compared to directly binned cross sections, which were generated according to Eqs.~(\ref{eq:direct_binning_diss})-(\ref{eq:direct_binning_elastic}). The same 10-bin system as before has been chosen to present the results. In Figs.~\ref{fig:cross_sections_comparison_15} to \ref{fig:cross_sections_comparison_diss}, a comparison is made for several state-to-state transitions involving the same examples bins as in Sec.~\ref{sec:urvc_arrhenius_fitting}. In these figures the small symbols delimiting the shaded areas beneath represent the directly binned QCT cross sections, while the solid lines show the analytically inverted ones. Several aspects are noteworthy. First, the directly binned cross sections exhibit noticeable statistical noise, a fact that was already mentioned in Sec.~\ref{sec:level_specific_processes} for the level-specific QCT cross sections. By contrast, in the cross sections obtained via the analytical inversion procedure this noise has been completely smoothed out. The second difference is that practically all directly binned cross sections show a marked decrease at energies between 15 and 20 eV. This was also noticed in Sec.~\ref{sec:level_specific_processes} when discussing cross sections involving individual level-to-level transitions. The gradual decrease in the cross sections implies that at high enough collision energies (and associated temperatures) the given process tends to become less and less frequent. The corresponding analytical cross sections are seen to decrease much more slowly (e.g. Fig.~\ref{fig:cross_sections_comparison_57}), or in some other cases keep increasing with collision energy (e.g. Fig.~\ref{fig:cross_sections_comparison_15}). Recall from the discussion surrounding Fig.~\ref{fig:example_cross_sections} that this behavior is controlled by the value of the exponent $\eta_R$ used within the functional form of Eq.~(\ref{eq:cross_section}). This parameter influences the curvature of the analytical cross sections, as well as their asymptotic behavior at high energies. Since $\eta_R$ is directly tied to the reaction-specific parameter $b_R$, the shape of the analytical curve is automatically determined by the analytical fit and cannot be easily changed without also affecting the associated reaction rate coefficients. In particular, Fig.~\ref{fig:cross_sections_comparison_15} shows the pair of cross sections for transition between bins 1 and 5. Fig.~\ref{fig:cross_sections_comparison_15}~(a) shows the excitation cross section $\bar{\sigma}_{1 \rightarrow 5}^E$, while Fig.~\ref{fig:cross_sections_comparison_15}~(b) represents the cross section for the corresponding deexcitation. Here, both analytical cross sections keep growing with increasing collision energy, because $\eta_{1 \rightarrow 5}^E = 1.095$ and thus slightly greater than one. Consistent with the rightmost plot in Fig.~\ref{fig:example_cross_sections}, near zero collision energy the analytical deexcitation cross section $\bar{\sigma}_{5 \rightarrow 1}^E$ quickly decreases to zero. As a counter-example, the opposite behavior is observed for the analytical cross sections for transitions between bins 5 and 7 shown in Fig.~\ref{fig:cross_sections_comparison_57}. In this case $\eta_{5 \rightarrow 7}^E = 0.732$, and one can see in Fig.~\ref{fig:cross_sections_comparison_57}~(b) how the deexcitation cross section obtained from direct binning tends to decrease as $E_t \rightarrow 0$, whereas its analytical counterpart diverges in this limit. For both plots shown in Fig.~\ref{fig:cross_sections_comparison_57}, as $E_t \rightarrow \infty$, the excitation- and deexcitation cross section obtained analytically tend towards zero, albeit much more slowly than their directly binned counterparts. Overall, for the 10-bin system discussed here, only about 16~\% of all excitation/deexcitation cross section pairs possess analytical fit parameters that lead to $\eta_R > 1$. The majority of these cross sections therefore behave as shown in the leftmost plot in Fig.~\ref{fig:example_cross_sections}. The third difference between the analytical- and directly binned cross sections was already alluded to in Sec.~\ref{sec:urvc_binning} and only affects the endothermic processes, i.e. those with an energy threshold $E_{R} > 0$. This threshold is captured well in the analytical curves, because it is explicitly built into the functional form of Eq.~(\ref{eq:cross_section}). By contrast, the directly binned cross sections do not necessarily follow this behavior. Taking a closer look at the excitation cross section $\bar{\sigma}_{1 \rightarrow 5}^E$ in Fig.~\ref{fig:cross_sections_comparison_15}, one can see that both the analytically inverted cross section (solid line) and the directly binned counterpart (black squares) become non-zero at approximately the same value of $E_t$. However, whereas the directly binned cross section appears to increase gradually with collision energy, the analytical curve has a distinct ``kink'' precisely at $\bar{E}_{1 \rightarrow 5}^E = 5.42 \, \mathrm{eV}$. By contrast, the problem of a ``smeared-out'' threshold is clearly observed in Fig.~\ref{fig:cross_sections_comparison_57}~(a), which shows the excitation cross section $\bar{\sigma}_{5 \rightarrow 7}^E$. Here, the directly binned cross section already becomes non-zero at values of $E_t$ which lie slightly below $\bar{E}_{5 \rightarrow 7}^E = 2.82 \, \mathrm{eV}$. Again, the corresponding analytical curve only becomes non-zero at precisely this threshold.


Following the internal energy exchange processes just discussed, Fig.~\ref{fig:elastic_cross_section_comparison} now shows a comparison of the intra-bin collision cross sections for bins $k$ = 1, 5 and 9. These are the same processes discussed in the context of Fig.~\ref{fig:rate_coefficient_bin_el}. As in previous comparisons, the directly binned QCT cross sections are represented by symbols above a shaded area, while the equivalent analytical curves making use of analytical fit parameters are plotted as solid lines. The first thing to notice is that the intra-bin collision cross sections are approximately one order of magnitude greater than those for most of the inelastic processes. This difference in magnitude is consistent with what we observed in Sec.~\ref{sec:urvc_arrhenius_fitting} where we compared the intra-bin collision rate coefficients with those for internal energy exchange and dissociation. Notice that the agreement between the directly binned QCT cross sections and their analytical counterparts in Fig.~\ref{fig:elastic_cross_section_comparison} is fairly good for collision energies up to about 12-15 eV. Beyond this value though, the QCT cross sections exhibit a rather sudden drop to zero, whereas the analytical curves decrease at a much slower rate. Since the collision pair's kinetic energy is conserved in intra-bin collisions, no energy threshold is required for this process, and the intra-bin collision cross sections remain non-zero even in the low-energy limit. However, in all three cases shown, a discrepancy between the QCT cross sections and their analytical counterparts can be observed. The analytical curves diverge as $E_t \rightarrow 0$, whereas the directly binned cross sections remain finite. This is due to the particular values of the parameter $\eta_R$ in the analytical form. With $\eta_{k\rightarrow k}^\mathrm{coll} < 1$, for all three curves shown, the analytical cross sections diverge in this limit. In addition to the cross sections obtained for the 10-bin system via the two methods, all three plots in Fig,~\ref{fig:elastic_cross_section_comparison} contain an additional curve. This dashed line represents the total cross section for the variable hard sphere (VHS/VSS) model~\cite{bird80a, koura91a}:
\begin{equation}
 \sigma_T^\mathrm{VHS} = \frac{\pi d_\mathrm{ref}^2}{\Gamma \left( 5/2 - \omega \right)} \left( \frac{\mathrm{k_B} T_\mathrm{ref}}{E_t} \right)^{\omega - 1/2} \label{eq:vhs_cross_section}
\end{equation}

Using the values $d_\mathrm{ref} = 2.88$~\AA, $\omega = 0.69$ and $T_\mathrm{ref} = 2880 \, \mathrm{K}$ mentioned in Sec.~\ref{sec:urvc_arrhenius_fitting} as the VHS tuning parameters in Eq.~(\ref{eq:vhs_cross_section}), one obtains the dashed curves in Fig.~\ref{fig:elastic_cross_section_comparison}~(a) to (c). As discussed previously, the precise values for $d_\mathrm{ref}$ and $\omega$ are usually obtained by fitting molecular transport properties predicted with a VHS cross section to more accurate theoretical models, or experimental data in the vicinity of $T_\mathrm{ref}$. This usually results in a good reproduction of the gas mixture's transport properties when performing DSMC simulations at, or near the reference temperature. The use of total cross section based on the VHS/VSS model has been common practice in the DSMC community for a long time. In the present case though, it is clear that the VHS/VSS parameters obtained in the standard manner significantly under-estimate the magnitude of the elastic cross section when compared to the intra-bin collision cross sections derived from the Ames database. This confirms that low-temperature transport-coefficient data are not enough to accurately predict the total cross section in high-temperature chemically reacting flows. Notice also from Fig.~\ref{fig:elastic_cross_section_comparison}, that the intra-bin collision cross sections extracted from the database are different for each of the bins considered. This dependence of the collision cross sections on the internal energy state of the collision partners has been observed by others~\cite{kim14a, zhu16a} and becomes relevant at the time of formulating a consistent state-to-state collision algorithm in the DSMC method.


Finally, Fig.~\ref{fig:cross_sections_comparison_diss} shows the comparisons between directly binned and analytically inverted cross sections for dissociation from bins 1, 5 and 9 out of the same 10-bin system. When examining the analytical cross sections in these figures, one can notice that for all three bins they continue to grow as $E_t \rightarrow \infty$. In fact, for the 10-bin system discussed here, all 10 bins possess analytical dissociation cross sections for which $\eta_k^{Df} > 1$. Thus, they all follow the same asymptotic behavior and continue to grow without bounds as $E_t$ tends towards infinity. Another aspect to notice is that bins 1 and 5 are composed only of bound levels, whereas bin 9 contains only predissociated levels. Recall from the discussion in Sec.~\ref{sec:urvc_arrhenius_fitting} that the energy threshold for dissociation from bound levels is positive, whereas it is zero for dissociation from predissociated levels. Thus, in Figs.~\ref{fig:cross_sections_comparison_diss}~(a) and \ref{fig:cross_sections_comparison_diss}~(b) the analytical dissociation cross sections for bins 1 and 5 both exhibit a clear threshold for dissociation at $E_{ 1}^{Df} = 2 E_\mathrm{N} - \bar{E}_1 = 8.86 \, \mathrm{eV}$ and $E_{ 5}^{Df} = 2 E_\mathrm{N} - \bar{E}_5 = 3.44 \, \mathrm{eV}$ respectively. By contrast, the cross section for endothermic dissociation from bin $k = 9$ shown in Fig.~\ref{fig:cross_sections_comparison_diss}~(c) has zero energy threshold and remains positive even as $E_t$ approaches zero. The effect this has on the relative magnitudes of the corresponding dissociation rate coefficients was exhibited in Fig.~\ref{fig:rate_coefficients_bin_diss}.

\section{Conclusions}
\label{sec:conclusions}

In this paper we have presented a methodology for extracting coarse-grained, state-resolved reaction cross sections from a database for the $\mathrm{N_2}$-$\mathrm{N}$ system, based on ab initio chemistry data developed at NASA Ames Research Center. In addition to this, we have modified the URVC bin model and adapted it for use within a coarse grain DSMC implementation. As a consequence of these changes one may re-interpret the much smaller number of uniform bins as a complete ``replacement'' of the full set of rovibrational levels from the original Ames database. These changes greatly simplify the DSMC implementation and allow to maintain consistency between the kinetic-scale and hydrodynamic-scale formulations of the URVC bin model. 

The major part of this paper consisted in developing a simple methodology for extracting the bin-specific reaction cross sections from the Ames database. We compared two alternative methods, which we call \emph{direct binning} and \emph{analytical inversion} respectively, with the help of a 10-bin example system. As can be seen in Sec.~\ref{sec:cross_sections_comparison}, the exact shapes of the cross sections obtained via the two methods in some instances exhibit noticeable differences. However, both sets of cross sections (directly binned and analytically inverted) roughly are of the same magnitude when evaluated at collision energies below 15-20~eV. This is fortunate, because it means that they will both predict similar macroscopic reaction rates, at least for the range of gas temperatures typical of atmospheric entry flows (roughly 300~K - 50000~K). The discrepancies encountered above 20~eV are less of a concern, if one considers that such high-energy collisions are much less frequent in the temperature range of interest. The somewhat convoluted procedure of first computing rate coefficients based on the full N3 database, subsequently reducing the mechanism by applying the URVC binning rules and finally inverting these coefficients back to obtain analytical expressions for the cross sections, has the effect of ``smoothing-out'' much of the noise contained in the original cross sections. The individual shape of each \emph{analytically inverted} cross section curve is controlled by only two Arrhenius-fit parameters, obtained from the corresponding rate coefficient. Compared to the tabulated cross sections obtained by \emph{direct binning}, this approach greatly reduces the amount of data necessary to specify the entire detailed-chemistry mechanism, which must be loaded into computer memory (i.e. RAM) at run-time. Furthermore, no interpolation is necessary to evaluate these analytical cross sections at any given $\mathrm{N_2}$-$\mathrm{N}$ pair's relative collision energy. Within our DSMC implementation, this makes them less costly in terms of CPU time compared to the directly binned ones. 

We are aware that compressing the full N3 reaction mechanism of the Ames database into a system consisting of a much smaller number of bins has an effect on the thermodynamic and chemical-kinetic properties of the gas mixture. In future work we will use the analytically inverted cross sections presented here in conjunction with a state-to-state implementation of the DSMC method. We will then study the effect of our model reduction on the internal energy excitation and dissociation processes in an adiabatic, constant-volume reactor, as well as across normal shock waves. 


\section*{Acknowledgments}

Research of T.E. Magin was sponsored by the European Research Council Starting Grant \#259354


\appendix

\section{Analytical fit parameters for URVC 10-bin system} \label{app:analytical_fit_parameters}

The analytical rate coefficients and cross sections plotted in Sec.~\ref{sec:urvc_arrhenius_fitting} and Sec.~\ref{sec:cross_sections_comparison} constitute only a small sample of the chemical-kinetic data for this 10-bin system. In Table~\ref{tab:plus3_all_fit_parameters} we list the fit parameters to generate the complete data-set. Recall that the analytical curves in Figs.~\ref{fig:rate_coefficients_bin_0001_0005}-\ref{fig:rate_coefficient_bin_el} and Figs.~\ref{fig:cross_sections_comparison_15}-\ref{fig:cross_sections_comparison_diss} for the \emph{forward} processes (i.e. dissociation, intra-bin collisions and excitation reactions), were generated using Eqs.~(\ref{eq:arrhenius_fit}) and (\ref{eq:cross_section}). \emph{Backward} deexcitation rate coefficients and cross sections were evaluated using Eqs.~(\ref{eq:urvc_bin_N3_deexcitation_rate_coefficient_evaluated}) and (\ref{eq:bin_deexcitation_cross_sections_substituted}) respectively. The bins' average energies and degeneracies are listed in Table~\ref{tab:7plus3_bins_equal_size_numbers}.

\begin{widetext}
\begin{singlespace*}

\begin{table}[H]
 \centering
 
 \caption{Analytical fit parameters for URVC 10-bin (7 bound:3 quasi-bound) system}
 \label{tab:plus3_all_fit_parameters}
  
 \begin{minipage}[t]{0.5\textwidth}
  \centering
  \subfloat[Dissociation]{\label{tab:7plus3_dissociation_fit_parameters}
  \begin{tabular}{c c c c c c c}
   $k$& & $A_k^{Df}$ [$\mathrm{cm^3/s}$] & & $b_k^{Df}$ & & $E_{ k}^{Df} / \mathrm{k_B}$ [$\mathrm{K}$] \\ \hline
    \\[-2ex]
    1 & & $4.37 \times 10^{-15}$ & & 0.927 & & $1.03\times 10^{5}$ \\
    2 & & $2.71 \times 10^{-15}$ & & 0.989 & & $8.80\times 10^{4}$ \\
    3 & & $1.54 \times 10^{-16}$ & & 1.294 & & $7.22\times 10^{4}$ \\
    4 & & $4.45 \times 10^{-17}$ & & 1.425 & & $5.61\times 10^{4}$ \\
    5 & & $4.47 \times 10^{-17}$ & & 1.451 & & $4.00\times 10^{4}$ \\
    6 & & $2.68 \times 10^{-16}$ & & 1.313 & & $2.38\times 10^{4}$ \\
    7 & & $1.69 \times 10^{-13}$ & & 0.763 & & $7.20\times 10^{3}$ \\
    8 & & $2.07 \times 10^{-13}$ & & 0.779 & & 0 \\
    9 & & $2.10 \times 10^{-13}$ & & 0.778 & & 0 \\
   10 & & $9.68 \times 10^{-13}$ & & 0.658 & & 0
  \end{tabular}}
 \end{minipage}~
 \begin{minipage}[t]{0.5\textwidth}
  \centering
  \subfloat[Intra-bin collisions]{\label{tab:7plus3_intra_bin_collision_fit_parameters}
  \begin{tabular}{c c c c c c c}
   $k$& & $A_{k \rightarrow k}^\mathrm{coll}$ [$\mathrm{cm^3/s}$] & & $b_{k \rightarrow k}^\mathrm{coll}$  & & $E_{k \rightarrow k}^\mathrm{coll} / \mathrm{k_B}$ [$\mathrm{K}$] \\ \hline
    \\[-2ex]
    1 & & $1.02 \times 10^{-10}$ & & 0.351 & & 0 \\
    2 & & $9.87 \times 10^{-11}$ & & 0.352 & & 0 \\
    3 & & $1.12 \times 10^{-10}$ & & 0.337 & & 0 \\
    4 & & $1.15 \times 10^{-10}$ & & 0.334 & & 0 \\
    5 & & $1.17 \times 10^{-10}$ & & 0.330 & & 0 \\
    6 & & $1.18 \times 10^{-10}$ & & 0.327 & & 0 \\
    7 & & $1.15 \times 10^{-10}$ & & 0.327 & & 0 \\
    8 & & $5.68 \times 10^{-11}$ & & 0.388 & & 0 \\
    9 & & $6.18 \times 10^{-11}$ & & 0.376 & & 0 \\
   10 & & $6.57 \times 10^{-11}$ & & 0.362 & & 0
  \end{tabular}}
 \end{minipage}
 
 \vspace{2mm}
 
 \begin{minipage}[t]{0.5\textwidth}
  \centering
  \subfloat[Excitation: pre-collision bins $k = 1 \dots 3$]{\label{tab:7plus3_excitation_fit_parameters_1to3}
  \begin{tabular}{c c c c c c c c c}
  $k$ & & $l$ & & $A_{k \rightarrow l}^{E}$ [$\mathrm{cm^3/s}$] & & $b_{k \rightarrow l}^{E}$  & & $E_{ k \rightarrow l}^{E} / \mathrm{k_B}$ [$\mathrm{K}$] \\ \hline
   \\[-2ex]
    1 & & 2 & & $6.01 \times 10^{-10}$ & & 0.015 & & $1.49 \times 10^{4}$ \\
    1 & & 3 & & $5.45 \times 10^{-13}$ & & 0.601 & & $3.07 \times 10^{4}$ \\
    1 & & 4 & & $2.43 \times 10^{-13}$ & & 0.674 & & $4.68 \times 10^{4}$ \\
    1 & & 5 & & $5.77 \times 10^{-13}$ & & 0.595 & & $6.29 \times 10^{4}$ \\
    1 & & 6 & & $1.88 \times 10^{-12}$ & & 0.488 & & $7.91 \times 10^{4}$ \\
    1 & & 7 & & $1.27 \times 10^{-11}$ & & 0.306 & & $9.57 \times 10^{4}$ \\
    1 & & 8 & & $1.54 \times 10^{-11}$ & & 0.227 & & $1.11 \times 10^{5}$ \\
    1 & & 9 & & $1.40 \times 10^{-12}$ & & 0.289 & & $1.31 \times 10^{5}$ \\
    1 & & 10 & & $2.49 \times 10^{-19}$ & & 1.487 & & $1.50 \times 10^{5}$ \\ \hline
    2 & & 3 & & $3.25 \times 10^{-10}$ & & 0.060 & & $1.58 \times 10^{4}$ \\
    2 & & 4 & & $8.24 \times 10^{-13}$ & & 0.563 & & $3.19 \times 10^{4}$ \\
    2 & & 5 & & $7.53 \times 10^{-13}$ & & 0.562 & & $4.80 \times 10^{4}$ \\
    2 & & 6 & & $2.08 \times 10^{-12}$ & & 0.468 & & $6.42 \times 10^{4}$ \\
    2 & & 7 & & $7.87 \times 10^{-12}$ & & 0.347 & & $8.08 \times 10^{4}$ \\
    2 & & 8 & & $1.19 \times 10^{-11}$ & & 0.271 & & $9.64 \times 10^{4}$ \\
    2 & & 9 & & $5.39 \times 10^{-12}$ & & 0.230 & & $1.17 \times 10^{5}$ \\
    2 & & 10 & & $2.77 \times 10^{-14}$ & & 0.491 & & $1.35 \times 10^{5}$ \\ \hline
    3 & & 4 & & $2.41 \times 10^{-10}$ & & 0.084 & & $1.61 \times 10^{4}$ \\
    3 & & 5 & & $1.35 \times 10^{-12}$ & & 0.520 & & $3.22 \times 10^{4}$ \\
    3 & & 6 & & $2.73 \times 10^{-12}$ & & 0.446 & & $4.84 \times 10^{4}$ \\
    3 & & 7 & & $9.37 \times 10^{-12}$ & & 0.336 & & $6.50 \times 10^{4}$ \\
    3 & & 8 & & $1.15 \times 10^{-11}$ & & 0.288 & & $8.06 \times 10^{4}$ \\
    3 & & 9 & & $1.39 \times 10^{-11}$ & & 0.183 & & $1.01 \times 10^{5}$ \\
    3 & & 10 & & $9.85 \times 10^{-14}$ & & 0.465 & & $1.19 \times 10^{5}$
  \end{tabular}}
 \end{minipage}~ 
 \begin{minipage}[t]{.5\textwidth}
  \centering
  \subfloat[Excitation: pre-collision bins $k = 4 \ldots 9$]{\label{tab:7plus3_excitation_fit_parameters_4to10}
  \begin{tabular}{c c c c c c c c c}
   $k$ & & $l$ & & $A_{k \rightarrow l}^{E}$ [$\mathrm{cm^3/s}$] & & $b_{k \rightarrow l}^{E}$  & & $E_{ k \rightarrow l}^{E} / \mathrm{k_B}$ [$\mathrm{K}$] \\ \hline
   \\[-2ex]
    4 & & 5 & & $1.66 \times 10^{-10}$ & & 0.121 & & $1.61 \times 10^{4}$ \\
    4 & & 6 & & $4.65 \times 10^{-12}$ & & 0.410 & & $3.23 \times 10^{4}$ \\
    4 & & 7 & & $1.35 \times 10^{-11}$ & & 0.308 & & $4.89 \times 10^{4}$ \\
    4 & & 8 & & $1.63 \times 10^{-11}$ & & 0.261 & & $6.45 \times 10^{4}$ \\
    4 & & 9 & & $3.24 \times 10^{-12}$ & & 0.333 & & $8.46 \times 10^{4}$ \\
    4 & & 10 & & $7.52 \times 10^{-13}$ & & 0.342 & & $1.03 \times 10^{5}$ \\ \hline
    5 & & 6 & & $2.32 \times 10^{-10}$ & & 0.090 & & $1.62 \times 10^{4}$ \\
    5 & & 7 & & $3.56 \times 10^{-11}$ & & 0.232 & & $3.28 \times 10^{4}$ \\
    5 & & 8 & & $4.38 \times 10^{-11}$ & & 0.185 & & $4.84 \times 10^{4}$ \\ 
    5 & & 9 & & $1.52 \times 10^{-11}$ & & 0.200 & & $6.85 \times 10^{4}$ \\
    5 & & 10 & & $3.08 \times 10^{-12}$ & & 0.236 & & $8.68 \times 10^{4}$ \\ \hline
    6 & & 7 & & $5.90 \times 10^{-10}$ & & 0.010 & & $1.66 \times 10^{4}$ \\
    6 & & 8 & & $1.99 \times 10^{-10}$ & & 0.061 & & $3.22 \times 10^{4}$ \\
    6 & & 9 & & $8.54 \times 10^{-11}$ & & 0.054 & & $5.23 \times 10^{4}$ \\
    6 & & 10 & & $1.56 \times 10^{-11}$ & & 0.103 & & $7.07 \times 10^{4}$ \\ \hline
    7 & & 8 & & $2.35 \times 10^{-09}$ & & -0.133 & & $1.56 \times 10^{4}$ \\
    7 & & 9 & & $2.05 \times 10^{-10}$ & & -0.025 & & $3.57 \times 10^{4}$ \\
    7 & & 10 & & $6.16 \times 10^{-11}$ & & -0.029 & & $5.41 \times 10^{4}$ \\ \hline
    8 & & 9 & & $4.21 \times 10^{-11}$ & & 0.154 & & $2.01 \times 10^{4}$ \\
    8 & & 10 & & $3.11 \times 10^{-12}$ & & 0.237 & & $3.84 \times 10^{4}$ \\ \hline
    9 & & 10 & & $6.07 \times 10^{-12}$ & & 0.246 & & $1.83 \times 10^{4}$
  \end{tabular}}
 \end{minipage}
\end{table}
 
\end{singlespace*}
\end{widetext}

\bibliography{rgd}{}
\bibliographystyle{unsrt}

\end{document}